\documentclass[twocolumn]{emulateapj}
\usepackage{float}

\begin{document}
\shorttitle{Low X-ray Luminosity Galaxy Clusters}
\shortauthors{Nilo Castell\'on et al.}
\slugcomment{Accepted for publication in The Astronomical Journal 03/25/2016}


\title{Low X-Ray Luminosity Galaxy Clusters: Main goals, \\
sample selection, photometric and spectroscopic observations}


\author{Jos\'e Luis Nilo Castell\'on\altaffilmark{1,2}, M. Victoria Alonso\altaffilmark{3,4}, Diego Garc\'ia Lambas\altaffilmark{3,4}, Carlos Valotto\altaffilmark{3,4}, 
Ana Laura O' Mill\altaffilmark{3,4}, H\'ector Cuevas\altaffilmark{1}, Eleazar R. Carrasco\altaffilmark{5}, Amelia Ram\'irez\altaffilmark{1} Jos\'e M. Astudillo\altaffilmark{6}, 
Felipe Ramos\altaffilmark{3}, Marcelo Jaque\altaffilmark{7}, Natalie Ulloa\altaffilmark{1} \& Yasna \'Ordenes\altaffilmark{8}}


\affil{$^{1}$ Departamento de F\'isica y Astronom\'ia, Facultad de Ciencias, Universidad de La Serena, Cisternas 1200, La Serena, Chile.}
\affil{$^{2}$ Direcci\'on de Investigaci\'on, Universidad de La Serena. Av. R. Bitr\'an Nachary 1305, La Serena, Chile.}
\affil{$^{3}$ Instituto de Astronom\'ia Te\'orica y Experimental, (IATE-CONICET), Laprida 922, C\'ordoba, Argentina.}
\affil{$^{4}$ Observatorio Astron\'omico de C\'ordoba, Universidad Nacional de C\'ordoba, Laprida 854, C\'ordoba, Argentina.}
\affil{$^{5}$ Gemini Observatory/AURA, Southern Operations Center, Casilla 603, La Serena, Chile.}
\affil{$^{6}$ Departamento de Ciencias F\'isicas, Universidad Andr\'es Bello, Av. Rep\'ublica 252, 837-0134 Santiago, Chile.}
\affil{$^{7}$ Instituto de Ciencias Astron\'omicas, de la Tierra y del Espacio (ICATE-CONICET), \\Av. Espa\~na Sur 1512, J5402DSP, San Juan, Argentina.\\}
\affil{$^{8}$ Argelander Institute f\"{u}r Astronomie der Universit\"{a}t Bonn, Auf dem H\"{u}gel 71, D-53121 Bonn, Germany.}

\begin{abstract}

We present the study of nineteen low X-ray luminosity galaxy clusters 
(L$_X \sim$ 0.5--45 $\times$ $10^{43}$ erg s$^{-1}$), selected from the 
 ROSAT Position Sensitive Proportional Counters (PSPC) 
Pointed Observations 
(Vikhlinin et al. 1998) and the revised version of Mullis et al. (2003) in 
the redshift range of 0.16 to 0.7.  

This is the introductory paper of a series presenting 
the sample selection, photometric and spectroscopic observations and data 
reduction.
Photometric data in different passbands were taken for eight galaxy clusters 
at Las Campanas Observatory; three clusters at  Cerro Tololo 
Interamerican Observatory; and eight clusters at the Gemini Observatory.
Spectroscopic data were collected for only four galaxy 
clusters using Gemini telescopes.

With the photometry, the galaxies were defined based on the star-galaxy 
separation taking into account photometric parameters. For each galaxy cluster,
the 
catalogues 
contain the PSF and aperture magnitudes of galaxies within the 
90\% completeness limit. They are used together with structural 
parameters to study the galaxy 
morphology and to estimate photometric 
redshifts.  With the spectroscopy, the derived galaxy velocity 
dispersion of our clusters ranged from 
507  km~s$^{-1}$ for [VMF98]022 to 775 km~s$^{-1}$ for [VMF98]097 with
signs of  substructure.  

Cluster membership has been extensively discussed taking into account 
spectroscopic and photometric redshift estimates. In this sense, members
are the 
galaxies within a projected radius of  0.75 Mpc from the X-ray emission
peak and with clustercentric velocities smaller than the cluster 
velocity dispersion or 
6000 km~s$^{-1}$, respectively.
These results will be used in forthcoming 
papers to study, among the main topics, the red cluster sequence, blue 
cloud and green populations; the galaxy luminosity function and cluster 
dynamics.

\end{abstract}





\section{Introduction}

The hierarchical model of structure formation predicts that the progenitors of the galaxy clusters are relatively small systems that are assembled together at higher redshifts.   Local cluster processes such as ram
pressure stripping and galaxy harassment play an important role 
in explaining the difference between
cluster and field galaxy populations at a fixed stellar mass (Berrier et al. 2009).
The study of galaxy systems in a variety of masses at different redshifts may 
give invaluable physical insights into galaxy evolution. 

The observed galaxy scaling relationships provide important tools for examining 
physical properties of galaxies and their systematics.  These 
relations
might be linked to the local galaxy density in rich
clusters (eg. Dressler 1980) or the galaxy morphology evolution (eg. Butcher \& 
Oemler 1984; Dressler \& Gunn 1992).  Their connection with different 
mechanisms such as galaxy 
collisions (Spitzer \& Baade, 1951) and interactions with 
intracluster gas (Gunn \& Gott 1972) are crucial to understand galaxy
formation and evolution.  When a galaxy cluster is assembled, 
the morphology, 
luminosity, mass, and mean stellar age of their member galaxies  are 
determined by these processes.   

Kodama et al. (1998) studied the Color-Magnitude Relation (CMR) in distant 
clusters and suggested the monolithic model for the formation of
early-type galaxies, but also mentioned 
other possibilities. In particular, an alternative scenario is 
the hierarchical merging model (Kauffmann \& Charlot 
1998; De Lucia et al. 2004).  The red cluster sequence found in optical CMRs 
(Gladders et al. 1998; De Lucia 
et al. 2004; Gilbank et al. 2008; Lerchster et al. 2011, hereafter RCS), which 
is dominated by 
non-star-forming, early-type galaxies (Zhu, Blanton, \& Moustakas
2010; Blanton \& Moustakas 2009) can be used to test these
models. Changes in the slope
and zero-point of this relation may be an indication of cluster evolution. 
In the CMRs, star-forming, 
late-type galaxies populate the ``blue cloud''. The presence 
of these two populations emerges as a bimodality in the 
color distribution (Baldry et al. 2004) as well as a ``green valley'' 
between them (see for instance, Mendez et al. 2011). 
The combination of deep images with multi-object spectrographs makes it
possible to explore additional evidence related to the processes responsible
for the observed properties of cluster members (Christlein \&
Zabludoff 2005) and to understand better their relationships with the 
environment (Finn et al. 2005).  

In the last twenty years, there was an increased interest in studying 
galaxy populations in clusters  due to the improvement in the observational 
facilities
that resulted in a large number of surveys.  The existence of the RCS; the 
blue galaxy population; and interactions as a function of redshift and 
environments, are among the main issues addressed by these surveys.  
We can mention: the ESO 
Distant Cluster Survey (EDisCS, White et al. 2005) in a wide range of mass,
with redshifts from 0.4 to
almost 1.0; the X-ray-luminous clusters from the MACS survey at z $\approx$ 
0.5 within 
a 1.2 Mpc diameter (Stott et al. 2007); the Observations of Redshift Evolution 
in Large-Scale Environments (ORELSE) Survey (Lubin et al. 2009), a systematic
search for structure around well-known clusters at redshifts of 0.6 $<$ z $<$ 1.3;
the galaxy populations in the core of a massive, X-ray luminous cluster 
(Strazzullo et al. 2010) at z=1.39; and the IMACS Cluster Building Survey (Oemler et al. 2013) 
to understand the large-scale environment surrounding rich intermediate redshift clusters of galaxies. 

On the other hand, regarding less massive clusters, 
Balogh et al. (2002) presented the first 
spectroscopic survey of low X-ray luminosity clusters (L$_X < 4 \times 10^{43}$ 
h$^{-2}$ erg s$^{-1}$ [0.1-2.4] keV) with Calar Alto spectroscopy
 and Hubble 
Space Telescope WFPC2 
imaging at 0.23 $<$ z $<$ 0.3. These clusters have 
Gaussian velocity distributions, with
velocity dispersions ($\sigma$) ranging from 350 to 850 kms$^{-1}$, consistent with
the local L$_X$-$\sigma$ relation.  The spectral and 
morphological properties of the galaxies in these systems were found similar to 
those in massive clusters at the same redshifts. 
Carrasco et al. (2007) analyzed the 
properties of the low luminosity X-ray cluster of galaxies RX J1117.4+0743 at 
z=0.485 based on optical and X-ray data finding a complex morphology
composed of at least two structures 
in velocity space. This cluster also presents an offset between 
the Bright Group Galaxy  and the X-ray emission.  
More recently, Connelly et al. (2012) have investigated systems detected in
both X-ray and the optical in the redshift range 0.12 $<$ z $<$ 0.79, 
obtaining a L$_X$--$\sigma$ scaling relation similar to observed in
 nearby groups. 

The X-ray properties of groups in 0.2 $<$ z $<$ 0.6
are the same as observed at lower redshifts (Mulchaey et al. 2006;
Jeltema et 
al. 2006).  In some cases, it was found that 
the X-ray emission was clearly peaked in the most luminous early-type galaxy.
There is also evidence that the central galaxy is composed of multiple 
luminous nuclei, suggesting that the brightest galaxy may still be undergoing 
major mergers.  At higher redshifts 
(0.85 $<$ z $<$ 1), Balogh et al. (2011) studied
the morphology of galaxies in six galaxy groups finding that they are 
dominated by red galaxies like lower redshift groups.  A few galaxies populate
the ``blue cloud'' and there is an important number of galaxies with 
intermediate colors, probable a transient population. 

Within this context, we aim at contributing to galaxy 
formation and evolution by analyzing a sample of low X-ray luminosity 
galaxy clusters 
at intermediate redshifts.  Our analysis can shed light on the properties
of these systems, in particular, the role of 
interactions  in the formation of galaxy clusters. 
In this paper, we  describe the cluster sample and the
data comprising photometric observations obtained at Las Campanas 
Observatory and Cerro Tololo 
Interamerican Observatory, and photometric and spectroscopic data
obtained at the Gemini Observatory.  These data will be used to study 
different galaxy populations in the 
clusters, as well as 
the galaxy luminosity function and cluster dynamics.
We have already published Nilo Castell\'on et al. (2014) and Gonzalez 
et al. (2015) 
on photometric galaxy properties and weak lensing analysis, respectively,
using part of this dataset. 
 The sample of low X-ray galaxy clusters is defined in detail in 
section $\S$2, the photometric
observations and the reduction procedures are given in section $\S$3 
including source detections and photometry, magnitude calibration, 
limiting magnitudes and completeness. In section $\S$4, 
the spectroscopic observations and data reductions are presented.
 In section $\S$5, the cluster membership assignment procedure is addressed
together with an outline of the photometric redshift estimates.
Finally in $\S$6, we present a brief comments on the project, the main results
already obtained and future plans. For all cosmology-dependent calculations, 
we have
assumed $\Omega_\Lambda$=0.7, $\Omega_m $=0.3 and $h$=0.7.

\section{Low X-ray Luminosity Cluster Sample}

Vikhlinin et al. (1998) presented the catalogue of 223 galaxy clusters 
based on the spatial extent of their X-ray emission, 
serendipitously detected in the ROSAT PSPC pointed observations with
photometric redshift estimates.
This catalogue of extended X-ray sources 
was revised by Mullis et al. (2003) using optical imaging and spectroscopy 
to classify 200 galaxy clusters, excluding 23 false detections. 
The spectroscopic cluster redshifts were derived by long-slit and multiobject
spectra with at least 2 or 3 concordant galaxy redshifts per cluster, 
always including the Bright Cluster Galaxy (BCG), and they entirely superseded
the photometric estimates of Vikhlinin et al. (1998).

\begin{deluxetable*}{lllrcccccc}
\tabletypesize{\scriptsize}
\tablewidth{0pc}
\tablecolumns{10}
\tablecaption{X-ray luminosity galaxy cluster sample.\label{table1}}
\tablehead{
\colhead{[VMF098]} & \colhead{ROSAT X-Ray} & \colhead{RA.} & \colhead{Decl.} & \colhead{$\delta$r} & \colhead{r$_{c}$} & \colhead{$\delta$r$_{c}$} & \colhead{L$_{X}$} & \colhead{$\delta$L$_{X}$} & \colhead{z} \\
\colhead{Id.} & \colhead{source} & \colhead{(J2000)} & \colhead{(J2000)} & \colhead{(arcsec)} & \colhead{(arcsec)} & \colhead{(arcsec)} & \colhead{ (10$^{43}$  erg s$^{-1}$)} & \colhead{(10$^{43}$  erg s$^{-1}$)} & \colhead{}}
\startdata
001  & RXJ0030.5+2618  &    00 30 33.2    & +26 18 19  & 13 & 31 & 3 & 26.1 & 3.2  & 0.500   \\
004  & RXJ0054.0-2823  &    00 54 02.8    & -28 23 58  & 16 & 37 & 6 & 4.2 & 0.6 & 0.292   \\
011  & RXJ0124.5+0400  &    01 24 35.1    & +04 00 49  & 20 & 31 & 14 & 3.4 & 1.0 & 0.316   \\
015  & RXJ0136.4-1811  &    01 36 24.2    & -18 11 59  & 15 & 21 & 8 & 1.4 & 0.3 & 0.251   \\
022  & RXJ0206.3+1511  &    02 06 23.4    & +15 11 16  & 14 & 53 & 10 & 3.6 & 0.7 & 0.248   \\
024  & RXJ0210.2-3932  &    02 10 13.8    & -39 32 51  & 11 & 22 & 10 & 0.6 & 0.1 & 0.168   \\
025  & RXJ0210.4-3929  &    02 10 25.6    & -39 29 47  & 14 & 28 & 9 & 0.8 & 0.2 & 0.165   \\
045  & RXJ0533.8-5746  &    05 33 53.2    & -57 46 52  & 37 & 81 & 28 & 8.7 & 2.4 & 0.297   \\
046  & RXJ0533.9-5809  &    05 33 55.9    & -58 09 16  & 30 & 53 & 20 & 1.6 & 0.5 & 0.198   \\
093  & RXJ1053.3+5720  &    10 53 18.4    & +57 20 47  &  8 & 12 & 3 & 1.4 & 0.2 & 0.340   \\
097  & RXJ1117.4+0743  &    11 17 26.1    & +07 43 35  & 12 & 18 & 7 & 6.4 & 1.7 & 0.477   \\
102  & RXJ1124.0-1700  &    11 24 03.8    & -17 00 11  & 22 & 34 & 19 & 8.1 & 2.5 & 0.407   \\
113  & RXJ1204.3-0350  &    12 04 22.9    & -03 50 55  & 14 & 26 & 6 & 2.7 & 0.4 & 0.261   \\
119  & RXJ1221.4+4918  &    12 21 24.5    & +49 18 13  & 18 & 34 & 8 & 42.7 & 9.5 & 0.700   \\
124  & RXJ1252.0-2920  &    12 52 05.4    & -29 20 46  & 13 & 46 & 11 & 3.4 & 0.7 & 0.188   \\
148  & RXJ1342.8+4028  &    13 42 49.1    & +40 28 11  & 16 & 15 & 6 & 16.2 & 4.4 & 0.699   \\
211  & RXJ2247.4+0337  &    22 47 29.1    & +03 37 13  & 20 & 46 & 17 &  4.1 & 1.1 & 0.200   \\
214  & RXJ2305.4-3546  &    23 05 26.2    & -35 46 01  & 15 & 55 & 14 & 2.8  & 0.6 & 0.201   \\
215  & RXJ2305.4-5130  &    23 05 26.6    & -51 30 30  & 17 & 21 & 10 & 0.7  & 0.2 & 0.194   \\
\enddata
\end{deluxetable*}

For the present work, we have selected systems with X-ray luminosities in the 
[0.5--2.0] keV energy band (rest frame), close to the detection limit 
of the ROSAT PSPC survey ranging from 0.1 to 
 50 $ \times 10^{43}$ erg s$^{-1}$.  These luminosities could be affected by
the presence of not removed point sources such as AGNs from the X-ray 
emission.  This effect could be more important at lower luminosities,
for instance groups containing an AGN could be wrongly included in the sample.
The redshift range of our selection is 
 0.16 to 0.70 where we have excluded well studied low redshift clusters as
well as the galaxy cluster [VMF98]061
at z $>$ 1 previously analyzed by Rosati et al. (1999).  Within these 
luminosity
and redshift limits, we have a sample of 140 galaxy clusters with low X-ray 
luminosities. After visual inspection, we have avoided those fields 
with bright stars and also those extended objects that would require
more than one image to cover the field with the 
available instruments and telescopes.  In this way, our 
studied sample 
corresponds to a random selection of 19 low X-ray galaxy clusters. 
Table~\ref{table1} presents a 
summary of the main characteristics of this studied galaxy cluster sample.
Columns (1) and (2) show the Vikhlinin et al. (1998) and the ROSAT X-Ray survey 
identifications.  The equatorial coordinates of the X-ray centroid and the position 
uncertainties are in columns (3) to (5); the cluster angular core radius 
(in arcsec) and the corresponding error are given
in columns (6) and (7); the X-ray luminosity in the [0.5--2.0] keV energy 
band and estimates of the lower bound
of their uncertainties are in columns (8) and (9); and the mean redshift in 
column (10).
Columns (3) to (7) are taken from Vikhlinin et al. (1998) while columns (8) and (10) are
from Mullis et al. (2003). The mean X-ray luminosity 
is  7.3 $\times$ $10^{43}$ erg s$^{-1}$, an intermediate/low luminosity when 
compared to $\sim 10^{42}$ erg s$^{-1}$ for groups with extended X-ray emission or 
larger values than $ \sim 5 \times 10^{44}$ erg s$^{-1}$ of rich clusters. 

\begin{figure*}
\begin{centering}
\includegraphics[width=160mm]{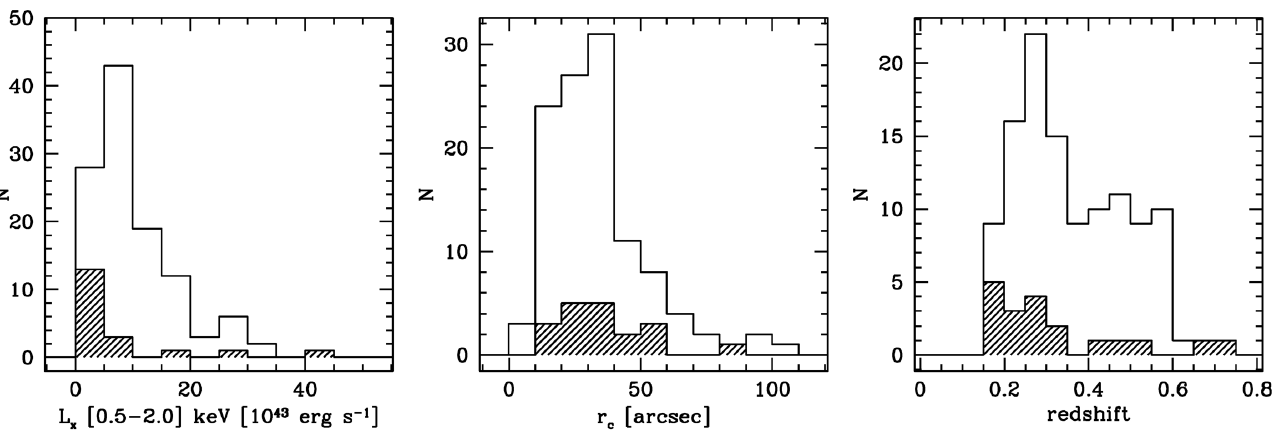}
\caption{The main galaxy cluster properties of the studied sample 
(shaded histograms) compared to the total 
sample of 140 galaxy clusters selected from Mullis et al. (2003). 
The X-ray luminosity, the angular core radius and the redshift distributions
are shown from the left to the right. }
\label{catalogue}
\end{centering}
\end{figure*}

Figure~\ref{catalogue} shows the distributions of X-ray luminosity,
angular core radius (in arcsec) and redshift of our
sample (shaded histograms) compared to the 
140 galaxy clusters selected from Mullis et al. (2003). It can be appreciated
that our sample of 19 galaxy clusters is 
biased towards low X-ray luminosities and lower redshifts
since we aim at studying this particular regime.  
The mean angular core radius was about 35 arcsec and most
of the galaxy clusters had values smaller than 60 arcsec.

\begin{figure}[!ht]
\begin{centering}
\includegraphics[width=80mm]{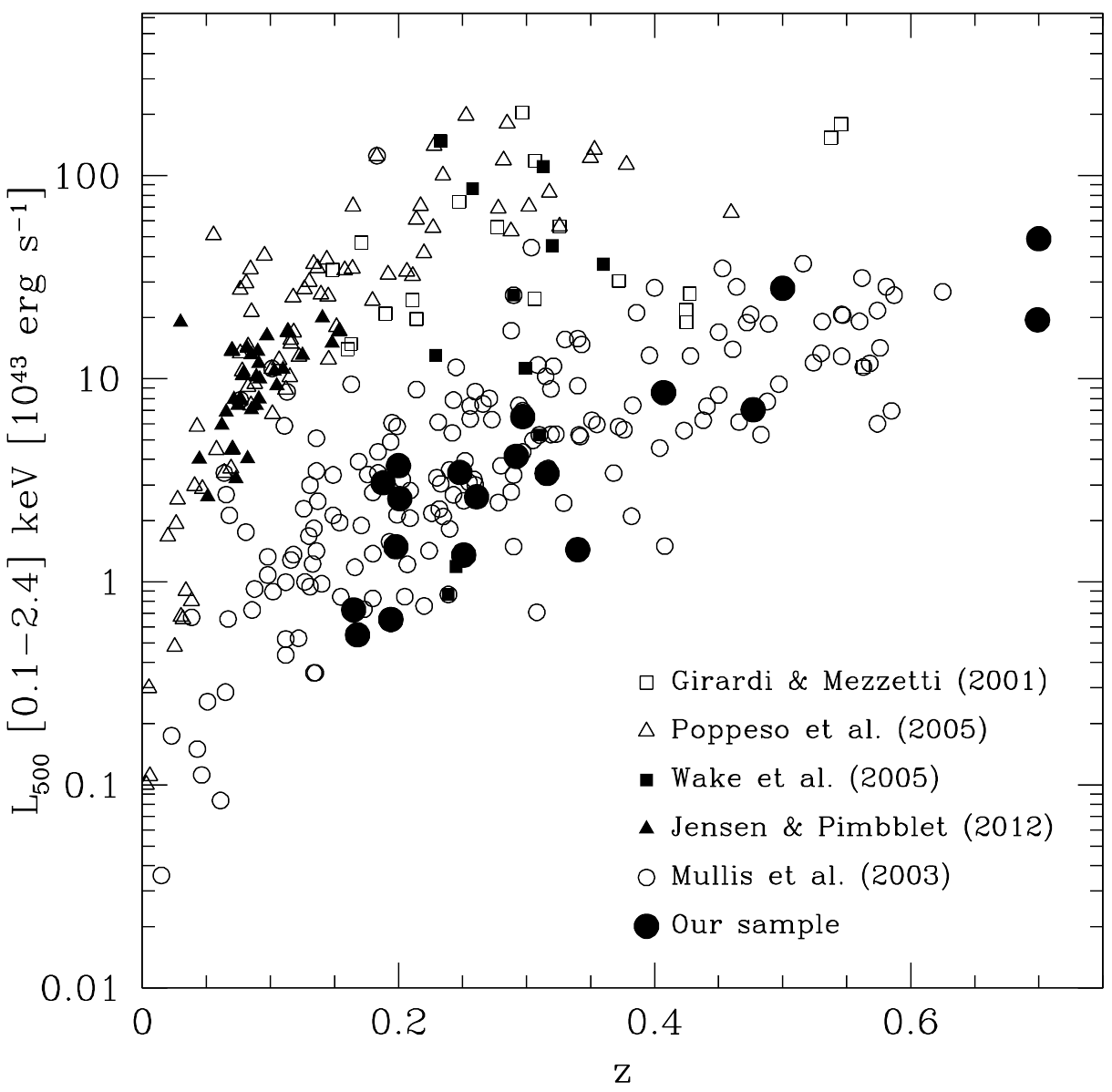}
 \caption{The X-ray luminosities (L$_{X}$ [0.5-2.0]) versus redshifts for 
different galaxy cluster samples.  The total sample of Mullis et al. (2003) 
and our cluster selection are represented by empty and filled circles,
respectively.  Other studies are also shown: 
 Girardi \& Mezzetti (2001,
 open squares); Popesso et al. (2005, open triangles); Wake et al. (2005,
 filled squares) and Jensen \& Pimbblet (2012, filled triangles).  The X-ray
luminosities are taken from the Piffaretti et al. (2011) compilation.}
 \label{sample}
\end{centering}
 \end{figure}

Figure~\ref{sample} shows 
the cluster X-ray luminosities (L$_{500}$ [0.1-2.4] keV) versus
redshifts of our galaxy cluster sample and the total sample from 
Mullis et al. (2003)  represented with
different circles.  We also include other works: 
Girardi \& Mezzetti (2001); Popesso et al. (2005); Wake et al. 
(2005) and Jensen \& Pimbblet (2012).  These studies 
analyze the
cluster membership and the photometric properties as 
galaxy colors and color-magnitude relations as our study.  All the X-ray 
luminosities are in the same system, extracted from Piffaretti et al. (2011)
which is the largest 
X-ray galaxy cluster compilation
based on publicly available ROSAT All Sky Survey data. In the redshift
range studied here, we have chosen the galaxy clusters with 
lower X-ray luminosities after discarding those mentioned above.  

The main goal of this work is to provide keys to understand the 
 cluster assembly and the
 morphological evolution of galaxies in low X-ray luminosity clusters.
  These 
systems thus provide us an interesting environment to explore the efficiency 
of mergers and ram pressure effects, 
that can be significantly different from those of rich galaxy clusters.
Our study is based on photometric observations
 of these low X-ray luminosity systems, where the high quality images also 
allowed us to construct a morphological catalogue to 
study galaxy morphologies and scaling
 relations.  We are particularly interested in a detailed study of the RCS
and an analysis of an eventual intermediate
green galaxy population between the galaxy blue cloud and the RCS 
in these low X-ray systems (Mendez et al. 2011).  For some galaxy clusters, we 
also carried out spectroscopic observations, which
allowed us to determine cluster membership and velocity dispersion 
estimates.

\section{Photometry}

In this section, we show the observations and the photometric procedures
adopted to obtain the galaxy properties of our cluster sample.

\subsection{Observations}

The galaxy clusters selected for this study have been observed using Las 
Campanas Observatory (LCO), Cerro Tololo 
Interamerican Observatory (CTIO) and Gemini Observatory.  

Eight galaxy clusters at z $<$ 0.32 were observed at LCO using 
the 2.5m du Pont telescope with the Wide Field Reimaging CCD Camera 
(TEK\#5 CCD) for direct imaging with a scale of 0.77 arcsec/pixel 
over a field of 25 arc-minute diameter using Chilean time allocation.  
The images were obtained in the $B, V, R,$ 
and $I$ Johnson-Cousins filters in nights with variable atmospheric conditions.
The seeing values were less than 1 arcsec in two galaxy clusters and
between 1.2 to 1.8 arcsecs in the remaining systems.  Six galaxy clusters
were observed in the 4 passbands while the clusters 
[VMF98]011 and [VMF98]211 only in the $B$ and $R$ passbands due to
poor atmospheric conditions.  

\begin{deluxetable*}{clllccccccc}
\tabletypesize{\scriptsize}
\tablewidth{0pc}
\tablecolumns{11}
\tablecaption{Photometric observations of the cluster sample.\label{table2}}
\tablehead{
\colhead{[VMF098]} & \colhead{Obs.} & \colhead{Obs.} & \colhead{Program} & \colhead{$B$} & \colhead{$V$} & \colhead{$R$} & \colhead{$I$} & \colhead{$g^{\prime}$} & \colhead{$r^{\prime}$} & \colhead{$i^{\prime}$} \\
\colhead{Id} & \colhead{Telescope} & \colhead{Date} & \colhead{Id.} & \colhead{} & \colhead{} & \colhead{} & \colhead{ } & \colhead{} & \colhead{} & \colhead{}}
\startdata
  		 001	     & GN  		& 10/06/2010 	& GN-2010B-Q-73  &-- 	           &--	            &-- 	     &-- & --	        &15$\times$300 &15$\times$150	      \\
  		 004	     & LCO  		& 09/18/2001 	& CNTAC 	& 5$\times$720 & 6$\times$600 &12$\times$600 &12$\times$600 & --            & --	         & --		  \\
 		 011	     & LCO  		& 09/22/2001 	& CNTAC 	& 5$\times$720 & -- &12$\times$600 & -- & --            & --	         & --		   \\
  		 015	     & LCO  		& 09/19/2001 	& CNTAC  	& 5$\times$300 & 5$\times$300 & 5$\times$300 & 5$\times$300 & --            & --	         & --		   \\
  		 022	     & GN  		& 05/21/2003   	& GN-2003B-Q-10   	&--              &--		    &--		     &--  & -- 	        & 4$\times$300 & 4$\times$150 	      \\
  		 024	     & LCO    		& 09/20/2001 	& CNTAC 		  & 5$\times$600 & 5$\times$600 & 5$\times$600 & 5$\times$600 & --	        & --	         & --	 \\
   		 025	     & LCO    		& 09/21/2001 	& CNTAC 		  & 5$\times$300 & 5$\times$300 & 5$\times$300 & 5$\times$300 & --	        & --	         & --	\\
  		 045	     & CTIO 		& 01/31/2001   	& DDT 			  &24$\times$900 &24$\times$900 &24$\times$600 &24$\times$600 & --	        & --	         & --	\\
  		 046	     & CTIO 		& 02/01/2001  	& DDT 			          &32$\times$700 &32$\times$700 &32$\times$500 &32$\times$500 & --	        & --             & -- \\
  		 093	     & GN  		& 06/24/2011 	& GN-2011A-Q-75 &--              &--		    &-- 	     &-- & --            & 5$\times$600 & 4$\times$150 	      \\
  		 097	     & GS  		& 06/24/2003   	& GS-2003A-SV-206			  &--		   &--		    &-- 	     &--&12$\times$600& 7$\times$900 & --	      \\
  		 102	     & GS  		& 07/03/2003  	& GS-2003A-SV-206  	           &--		   &--		    &-- 	     &--	   & --	        & 5$\times$600 & --     \\
  		 113	     & CTIO 		& 01/31/2001   	& DDT 				  &32$\times$900 &32$\times$900 &32$\times$600 &32$\times$600 & --	        & --	         & -- \\
 		 119	     & GN  		& 03/13/2011 	& GN-2011A-Q-75 	     &--  	 &--		   &--		    &--  & --	        & 7$\times$190 & 4$\times$120    \\
  		 124	     & GS    		& 06/24/2003  	& GS-2003A-SV-206  			  	     &--  	   &--		   &--		    &--	 &5$\times$300 & 5$\times$600 & --  \\
 		 148	     & GN  		& 02/28/2011 	& GN-2011A-Q-75  &-- 	           &--	            &-- 	     &-- & --	        & 7$\times$190 & 5$\times$120	      \\
  		 211	     & LCO  		& 09/22/2001 	& CNTAC  		  & 5$\times$600 & -- & 5$\times$600 & --  &--		   &--		    &-- \\
  		 214	     & LCO  		& 09/18/2001 	& CNTAC  		  & 5$\times$600 & 5$\times$600 & 5$\times$600 & 5$\times$600 & --	        & --	         & --	\\
  		 215	     & LCO    		& 09/20/2001 	& CNTAC 	         &12$\times$600 &12$\times$600 &12$\times$600 &12$\times$600 & --	        & --	         & -- \\
\enddata
\end{deluxetable*}

Three galaxy clusters at 0.19 $<$ z $<$ 0.30 were observed at CTIO using the 
Victor Blanco 4m telescope and the MOSAIC-II camera, which is an array of eight 
2048$\times$4096 SITe CCDs, with a scale of 0.27 arcsec/pixel, giving a total field of 
view (FOV) of 36$\times$36 arcmin.  The images were taken 
in the $B, V, R,$ and $I$ Johnson-Cousins passbands using the Director 
Discretionary Time.  The 
median seeing of the observations were about 0.85 arcsec. 

Finally, eight galaxy clusters with 0.18 $<$ z $<$ 0.70 were obtained using the
8m Gemini North (GN) and South (GS) telescopes in the $g^\prime$, $r^\prime$ and
$i^\prime$ passbands. The cluster [VMF98]102 had only observations in  
the $r^\prime$ band and it is included in the sample as spectroscopic 
observations were also made.  The Gemini Multi-Object Spectrograph 
(Hook et al. 2004, hereafter GMOS) was used in the image mode during the 
system verification process (SVP) and specific 
programmes using Argentinian time allocation, with the detector being
an array of three EEV CCDs of 2048$\times$4608 pixels. Using a 2$\times$2 
binning, the pixel scale is 0.1454 arcsec/pixel which corresponds to a
FOV of 5.5$\times$5.5 arcmin$^2$ in the sky. All images were observed 
under photometric conditions with excellent seeing values, mean estimates being 
less than 0.8 arcsec. 

Table~\ref{table2} shows cluster identifications and a summary of the photometric observations, 
including observatory, observation date, programme identification and number of 
exposures per filter with the individual exposure time given in seconds.  
The galaxy clusters and Landolt (1992) standard stars were observed with 
different filters, depending on the observational run.  

   \begin{figure*}
   \centering
	\includegraphics[width=0.2\hsize]{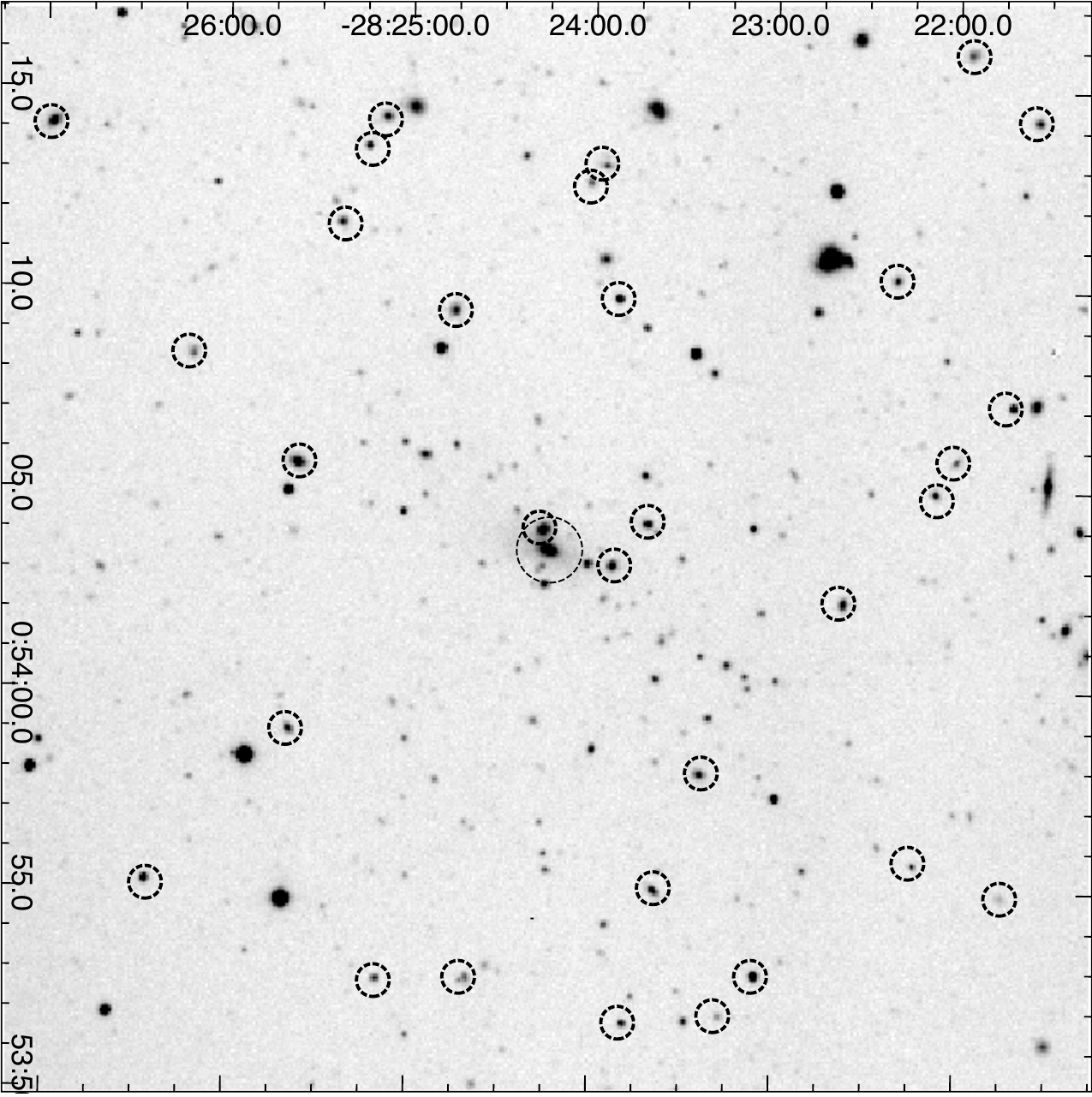}%
	\includegraphics[width=0.2\hsize]{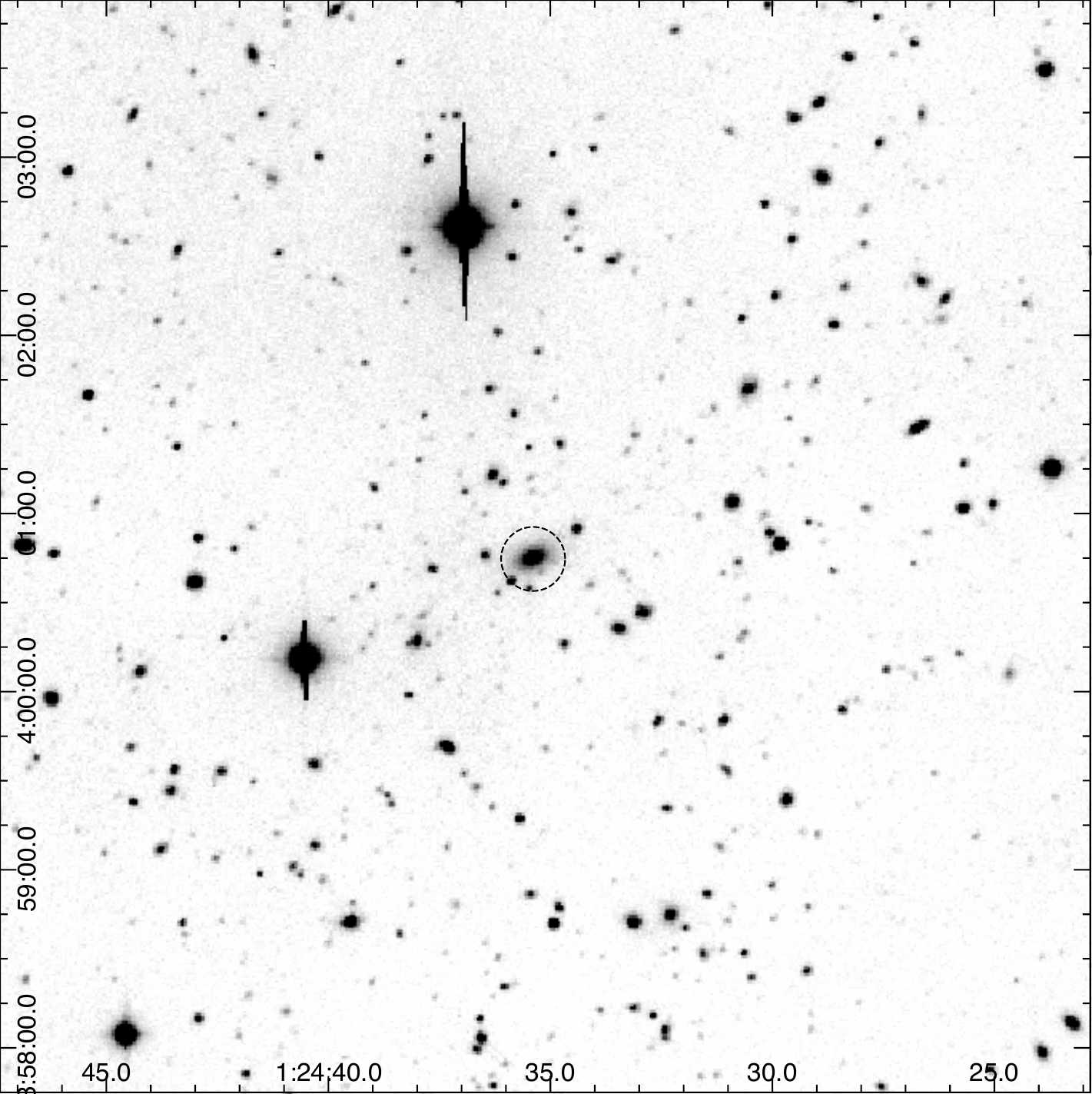}%
   	\includegraphics[width=0.2\hsize]{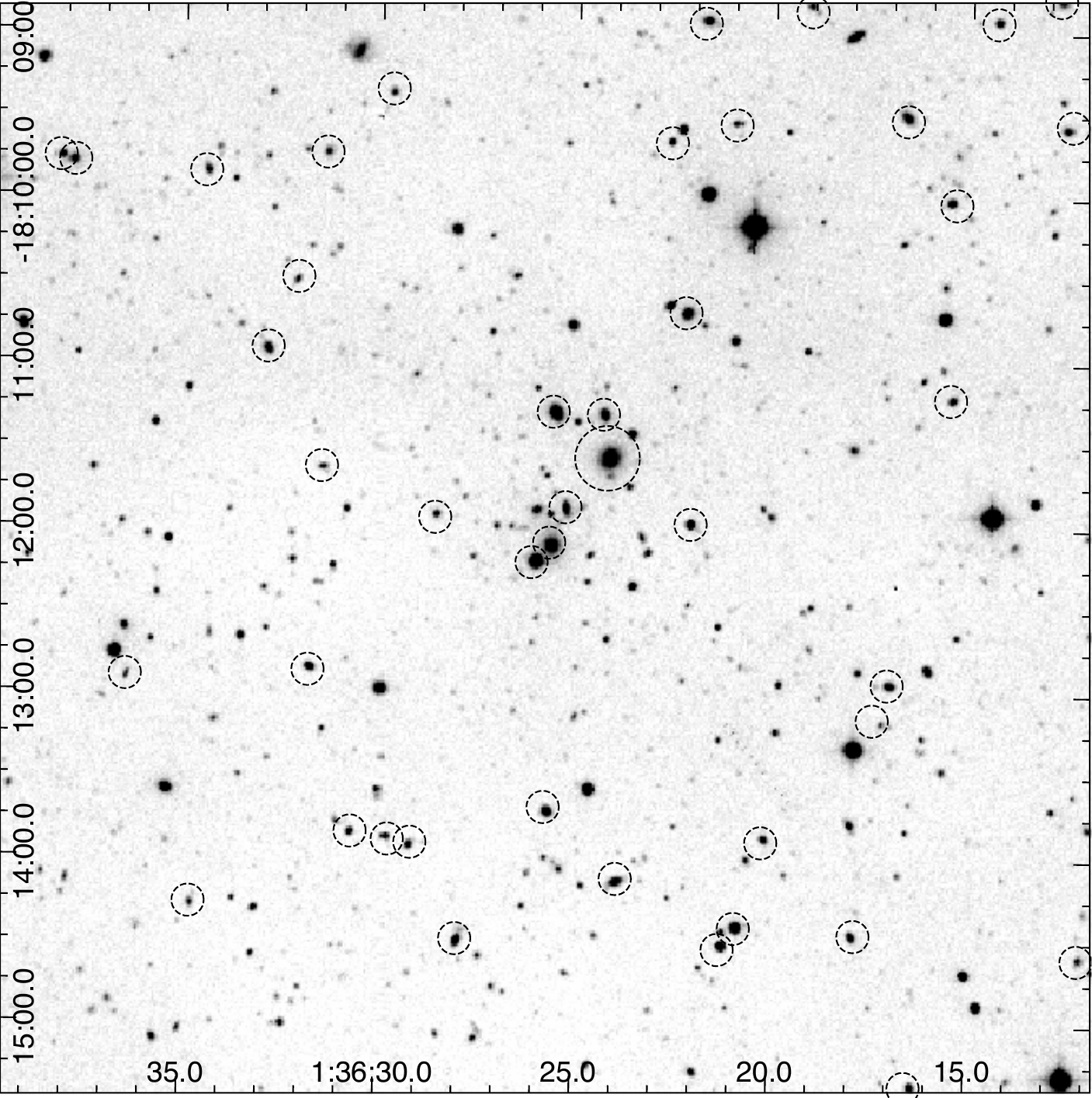}\\
	\includegraphics[width=0.2\hsize]{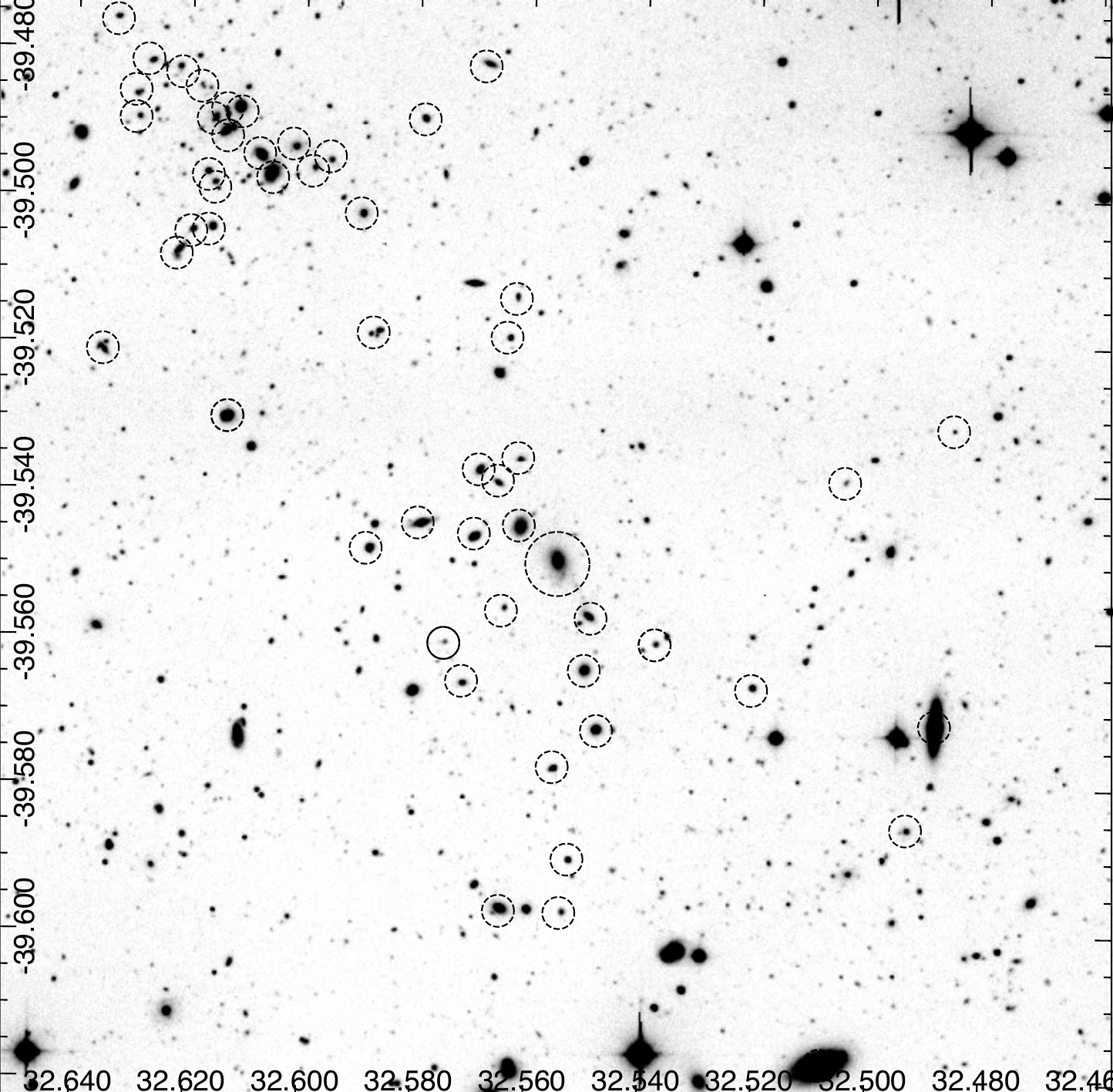}%
\includegraphics[width=0.2\hsize]{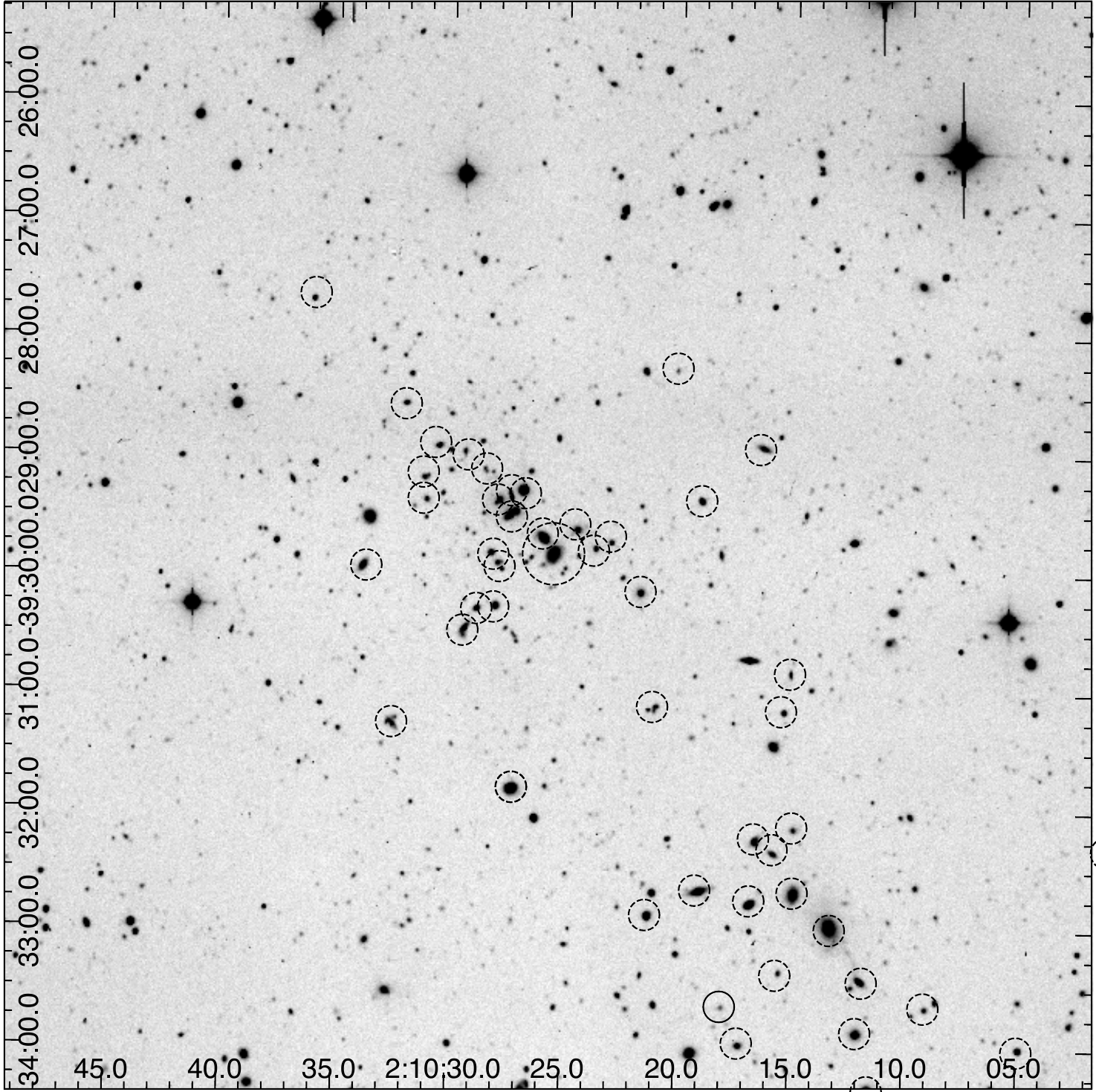}%
\includegraphics[width=0.2\hsize]{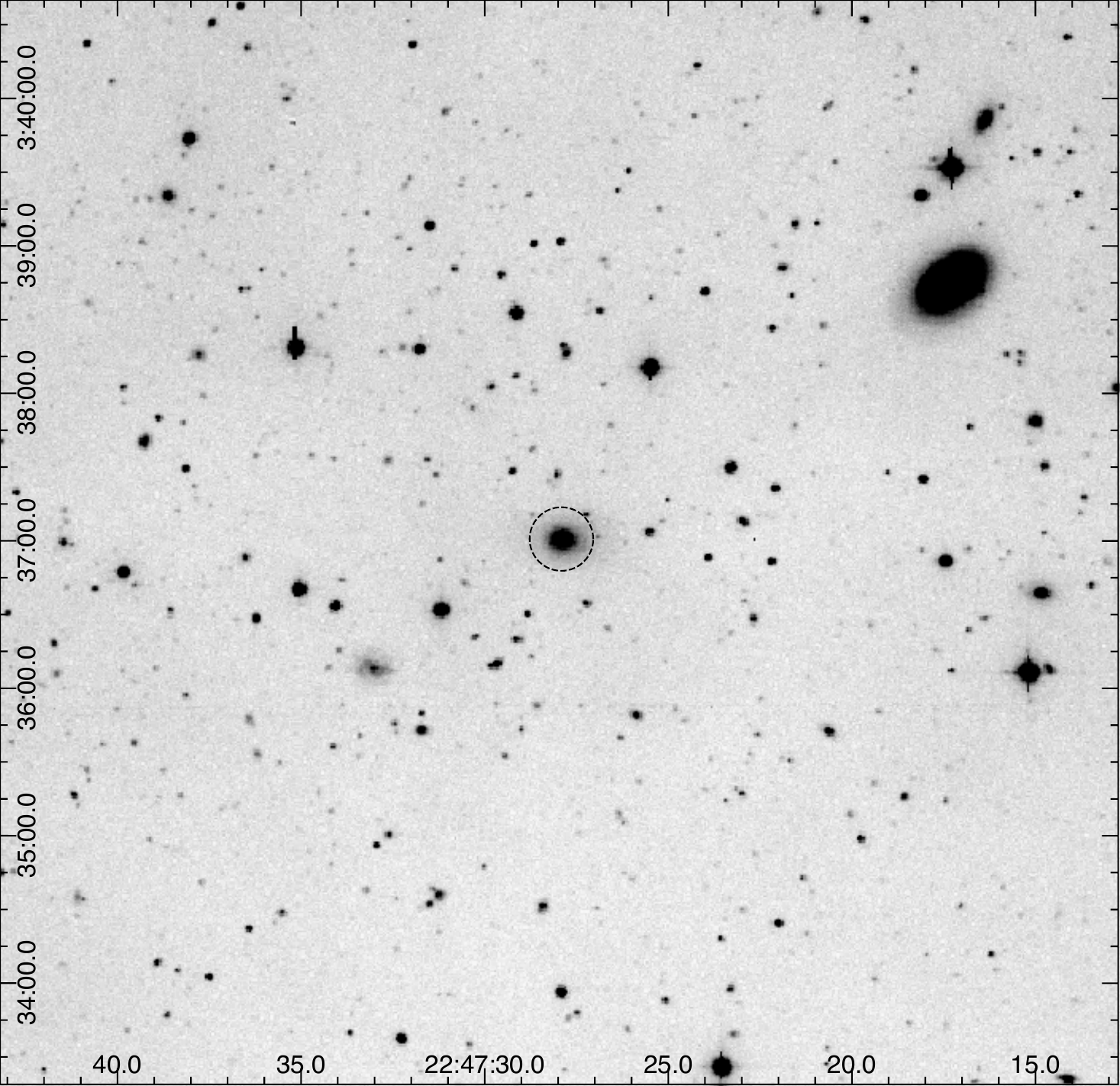}\\
   	\includegraphics[width=0.2\hsize]{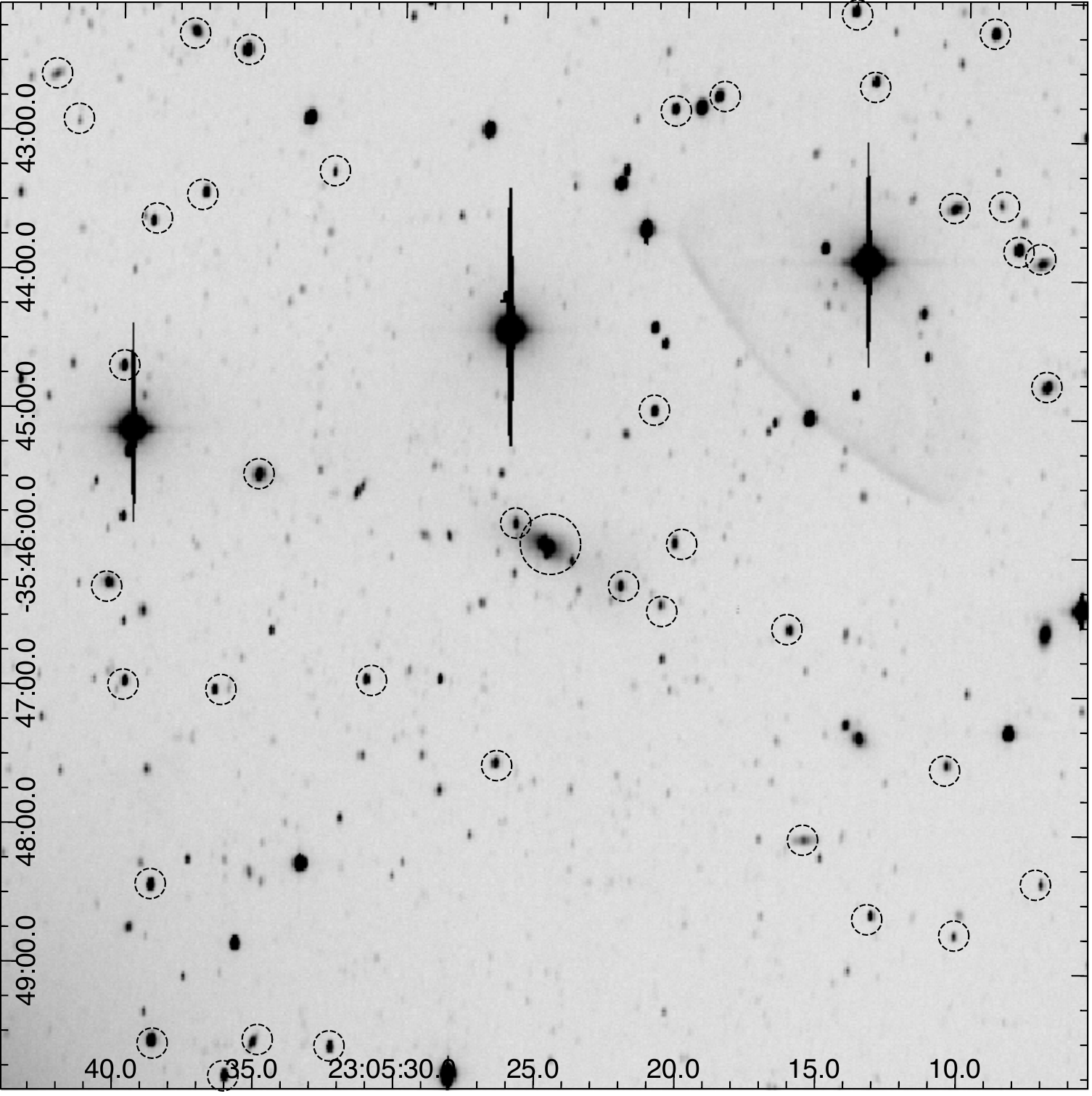}%
   	\includegraphics[width=0.2\hsize]{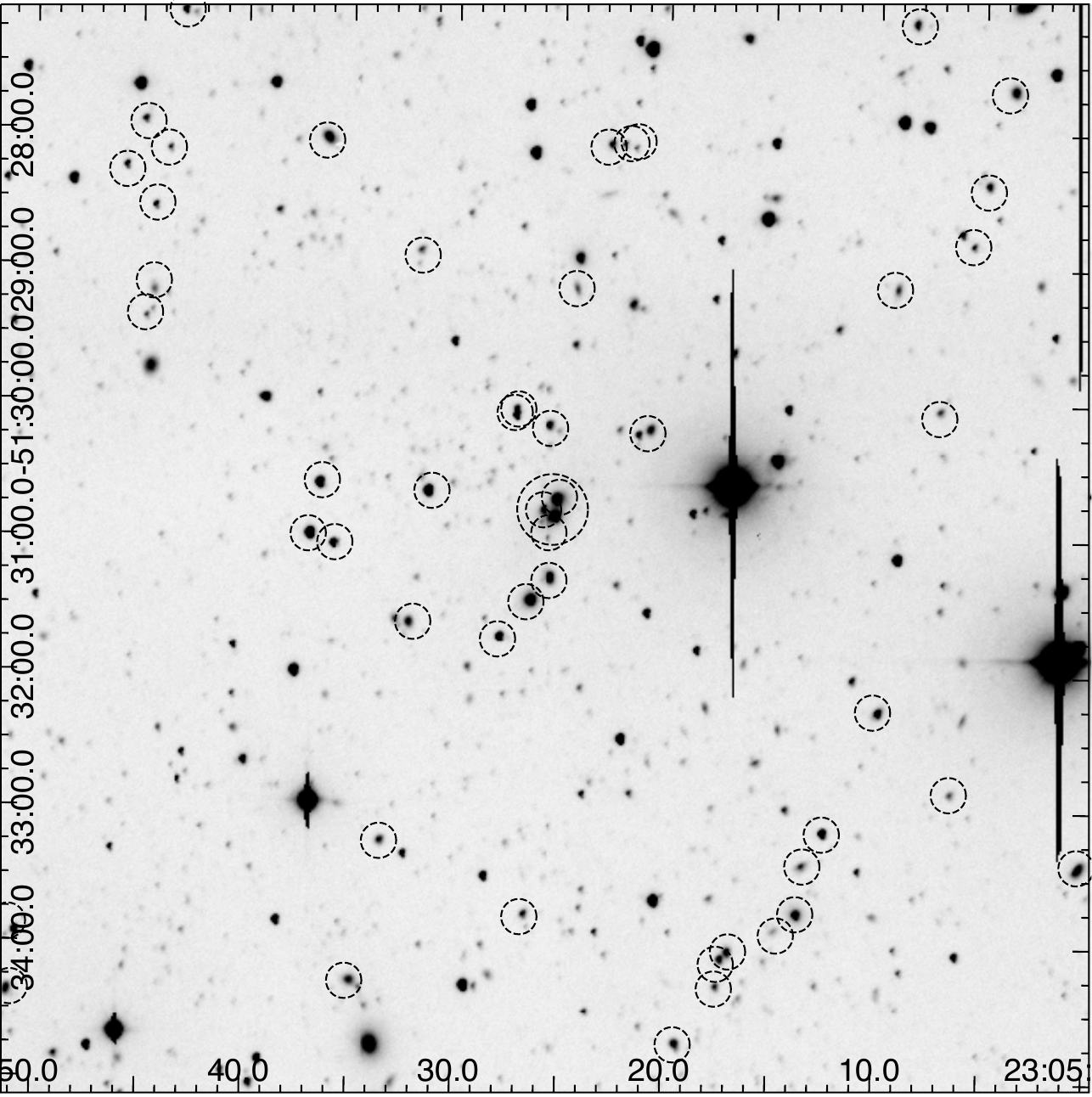} 
	\caption{R images of clusters observed with du Pont 2.5m telescope at LCO: [VMF98]004, [VMF98]011, [VMF98]015, [VMF98]024, [VMF98]025, 
[VMF98]211, [VMF98]214 and [VMF98]215 (from upper left to lower right panels). 
  The images are of 1.5 Mpc side. North is up and East is to the left.}
         \label{fov1}
   \end{figure*}

   \begin{figure*}
   \centering
        \includegraphics[width=0.2\hsize]{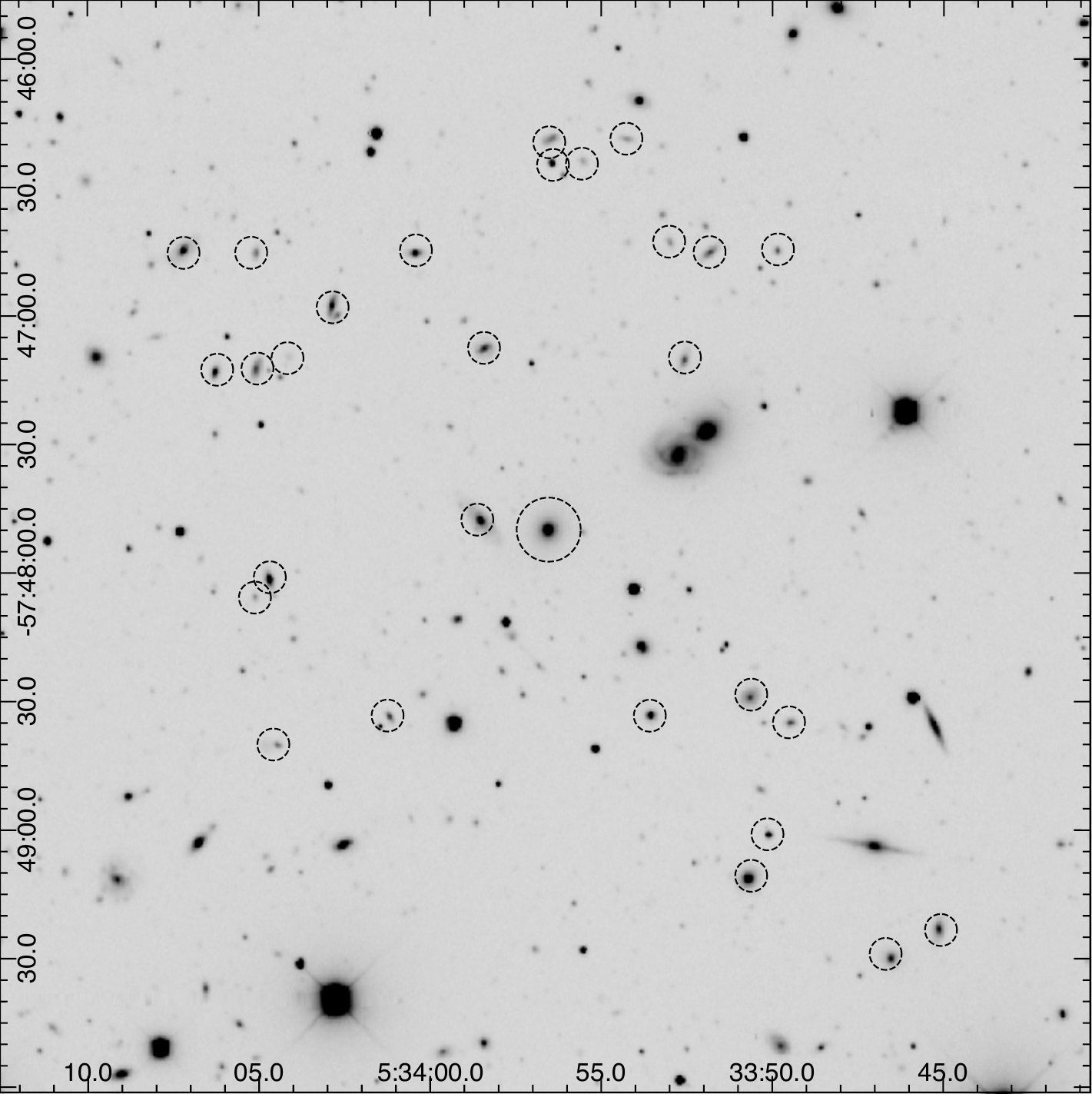}%
\includegraphics[width=0.2\hsize]{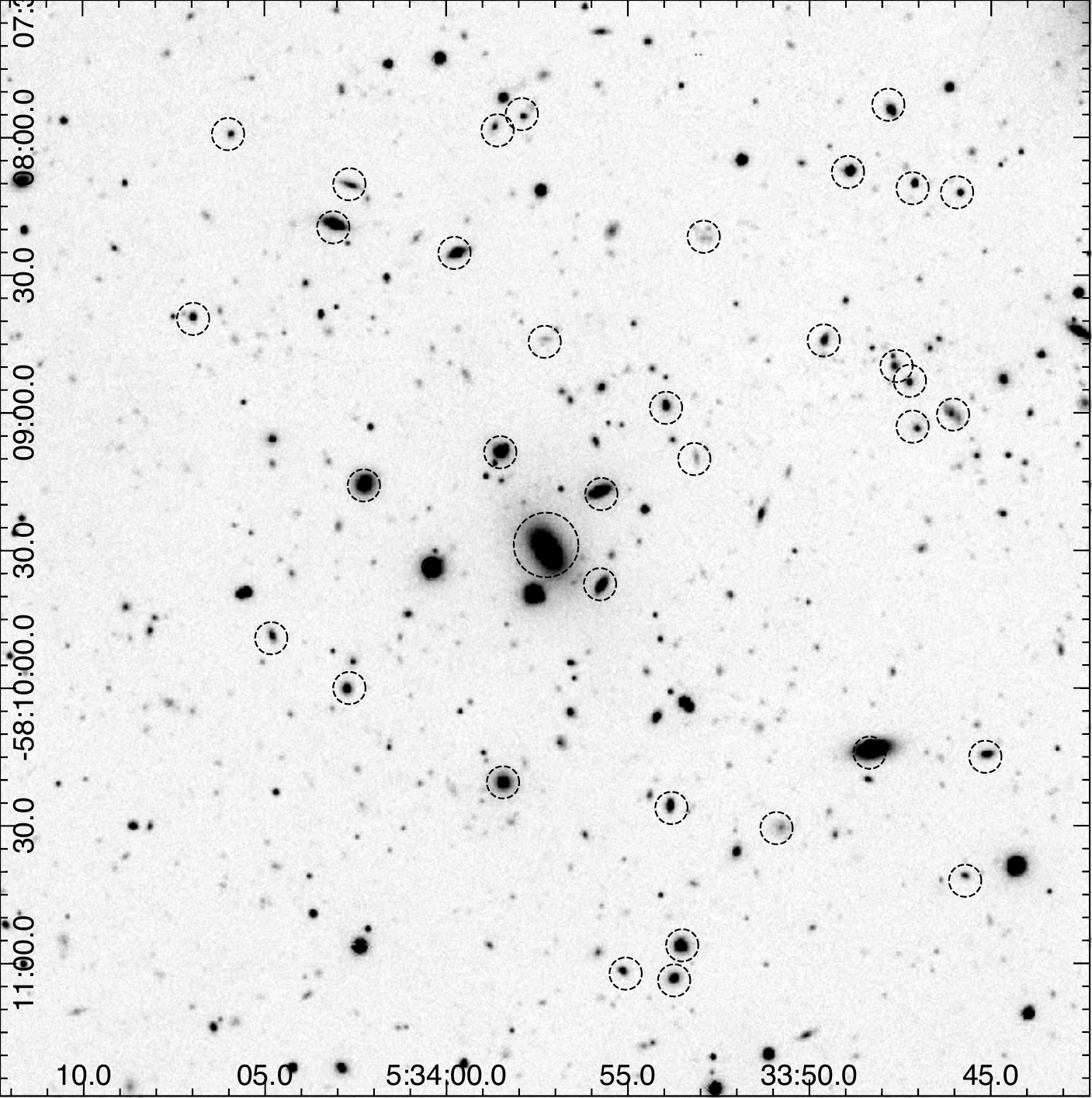}%
\includegraphics[width=0.2\hsize]{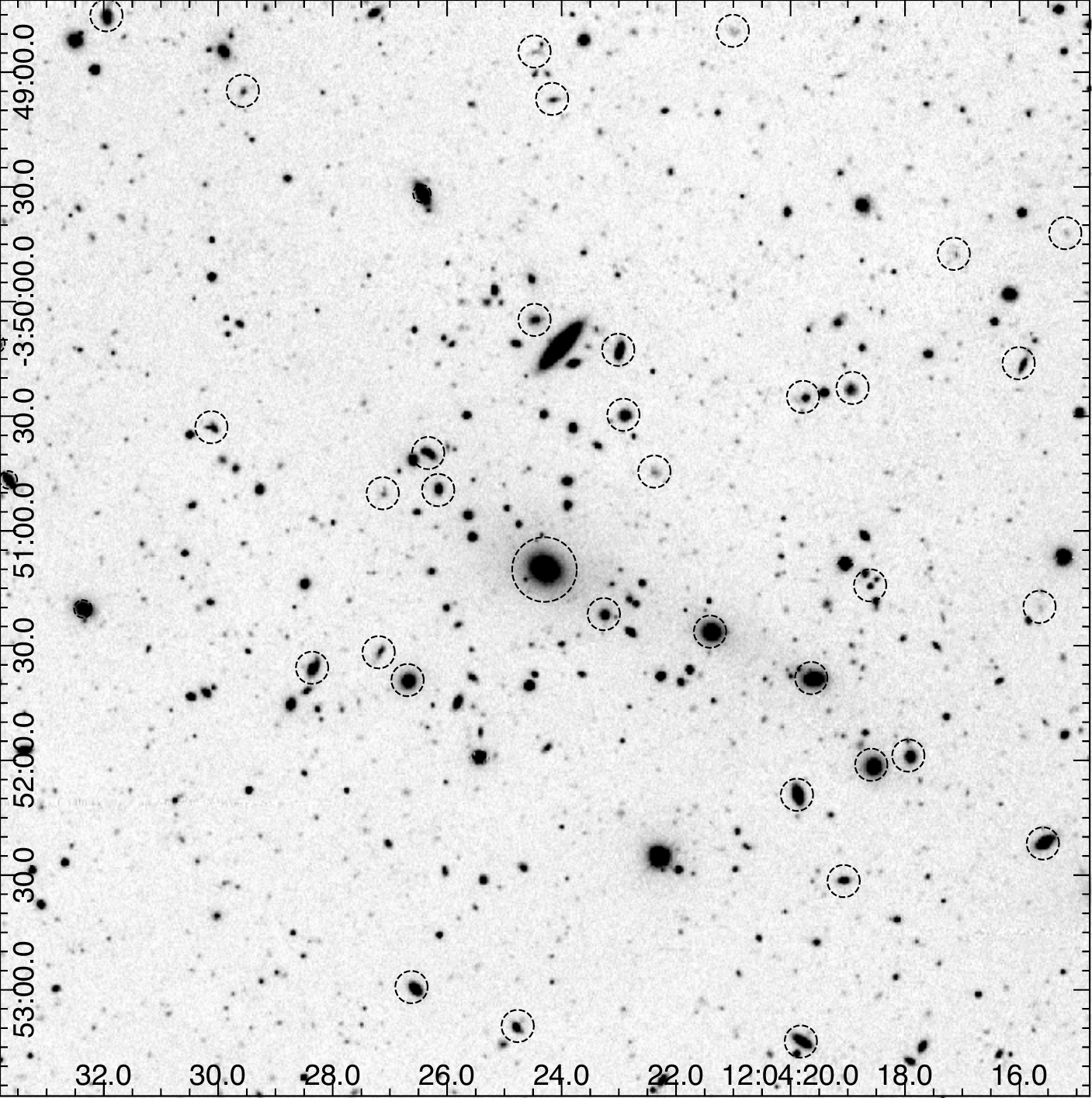}
	\caption{R images of the clusters observed with the CTIO Victor Blanco 4m telescope: [VMF98]045, [VMF98]046 and 
[VMF98]113 (from left to right). The images are of 1.5 Mpc side.  North is up and East is to the left. }
         \label{fov2}
   \end{figure*}
   
   \begin{figure*}
   \centering
        \includegraphics[width=0.2\hsize]{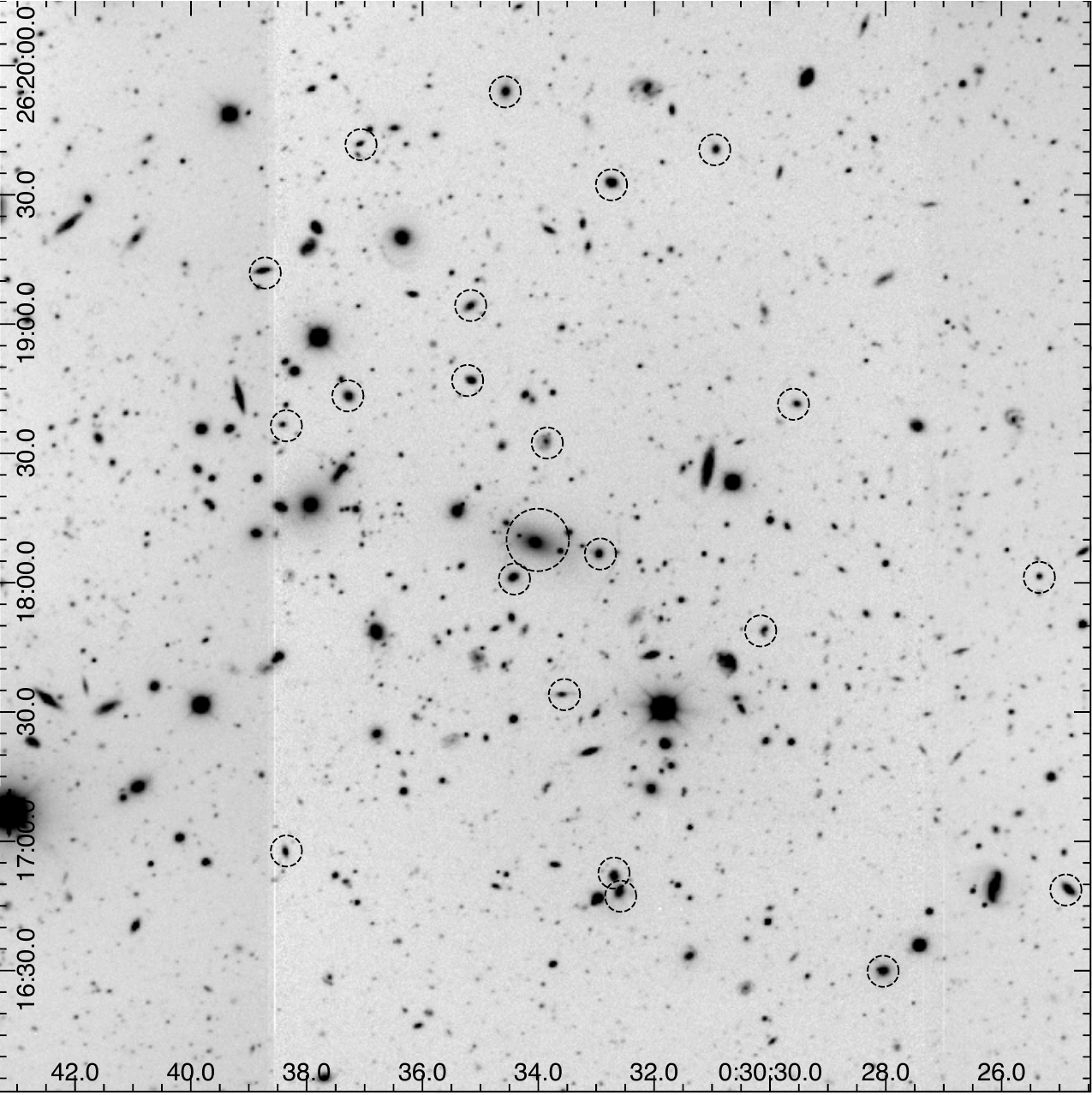}%
	\includegraphics[width=0.2\hsize]{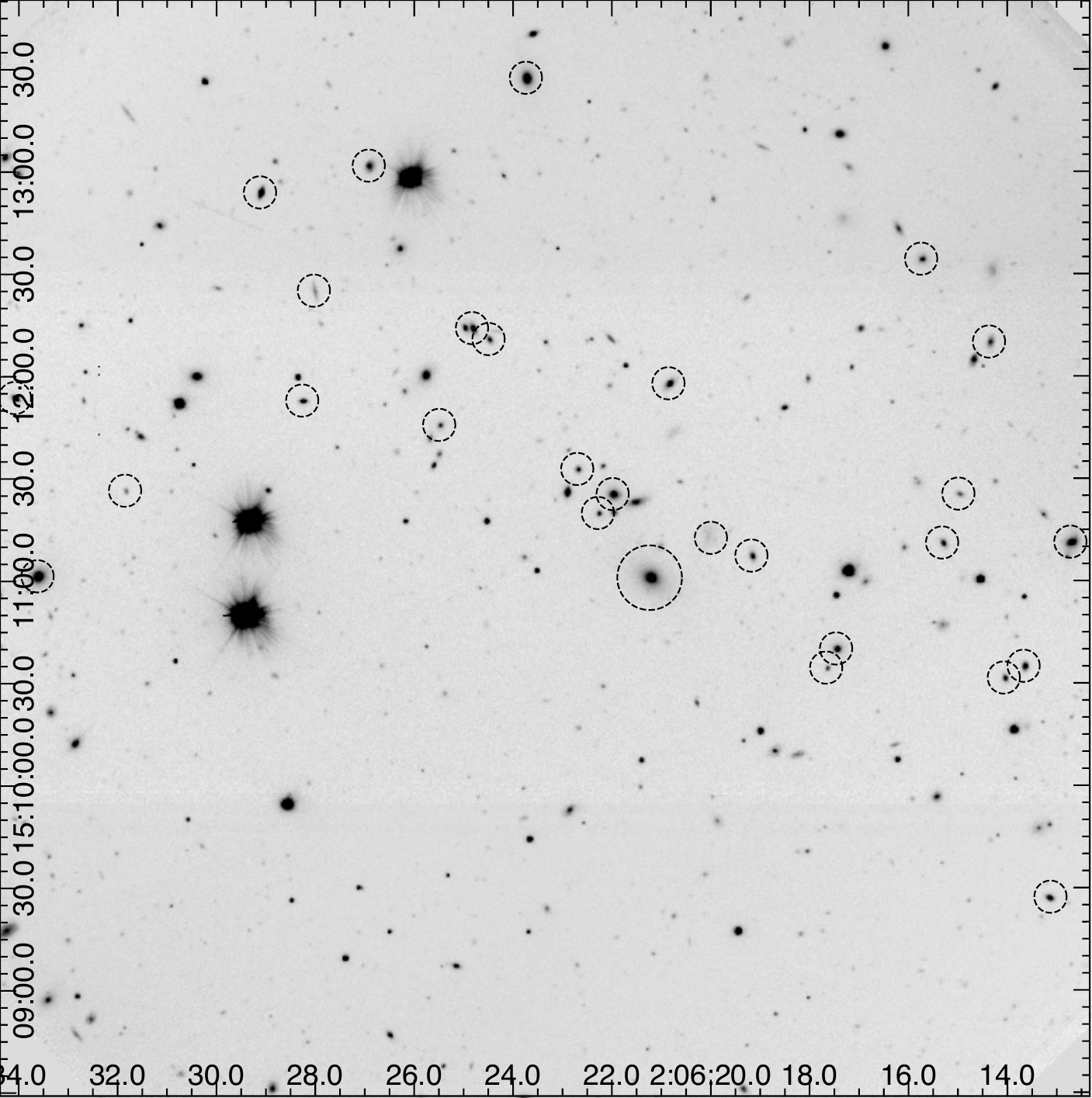}%
\includegraphics[width=0.2\hsize]{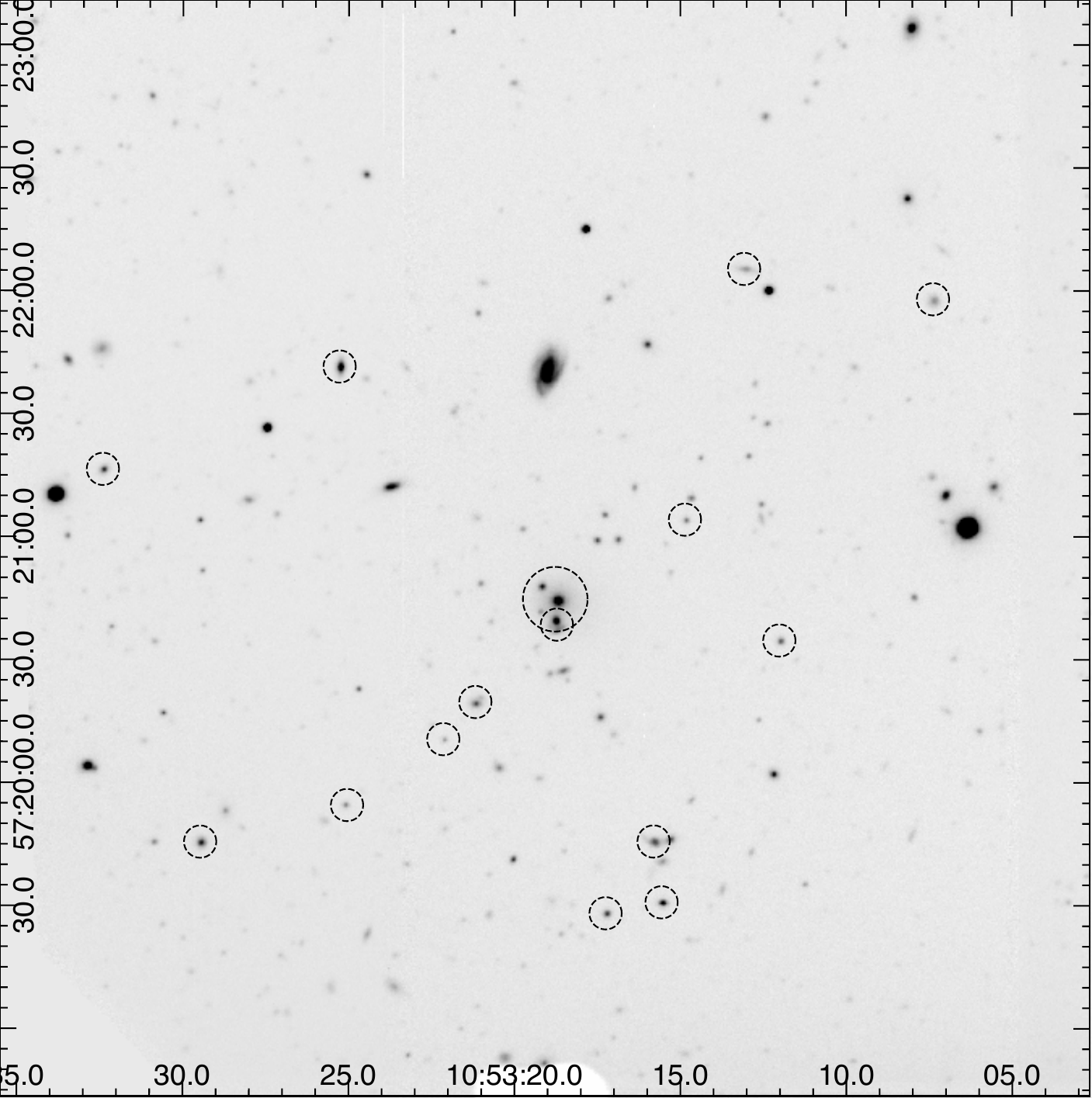}\\
\includegraphics[width=0.2\hsize]{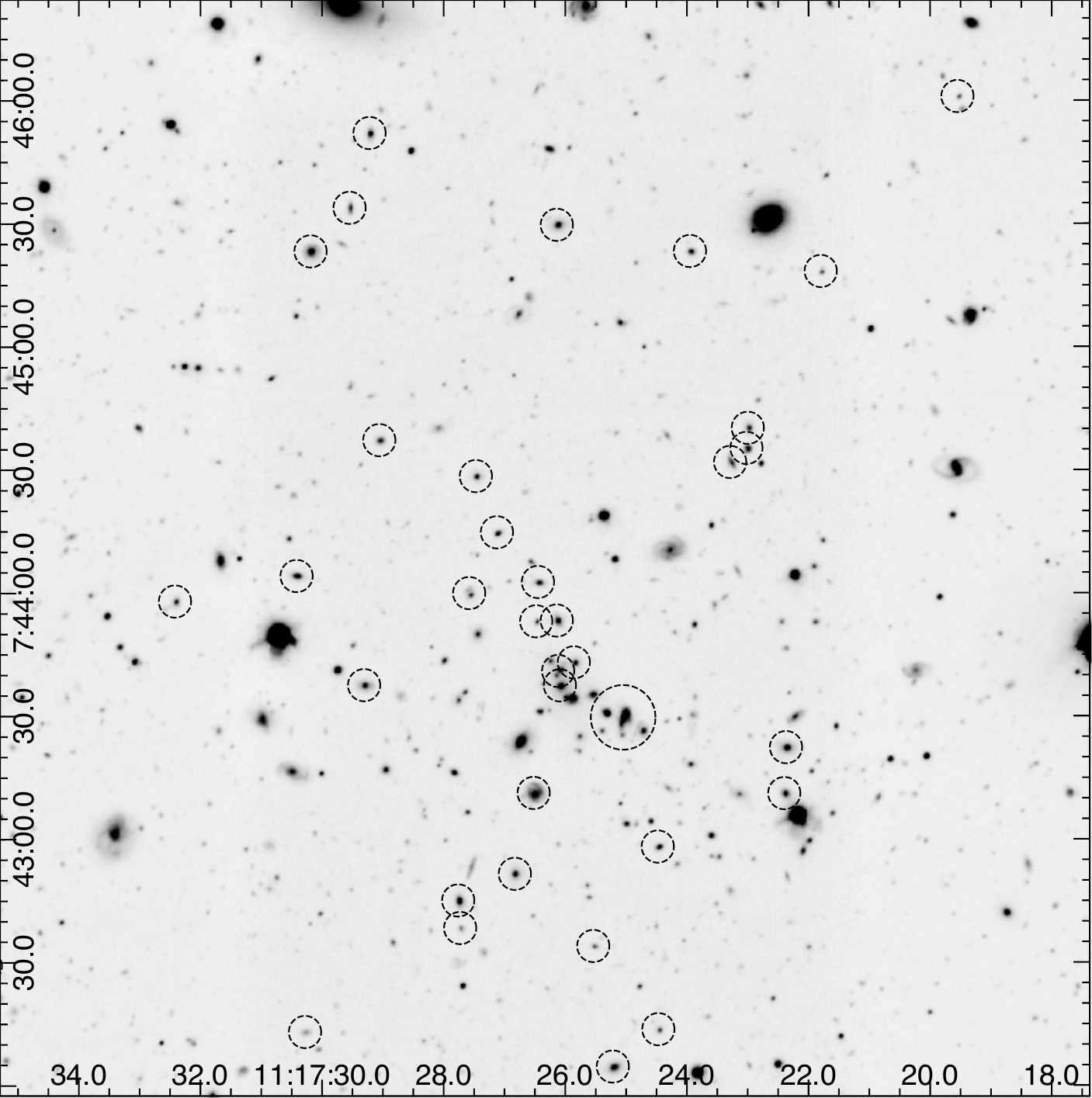}%
\includegraphics[width=0.2\hsize]{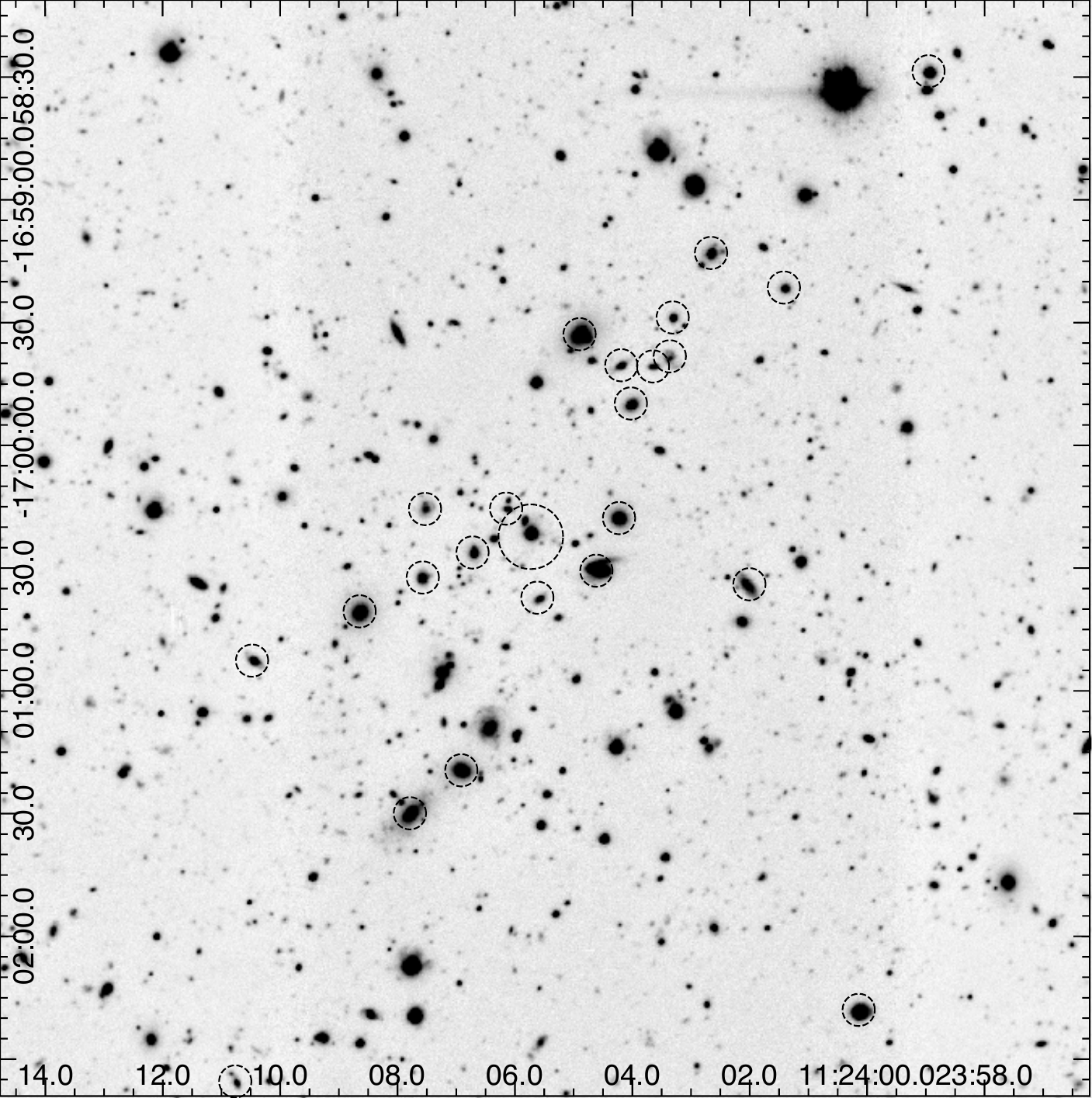}%
       \includegraphics[width=0.2\hsize]{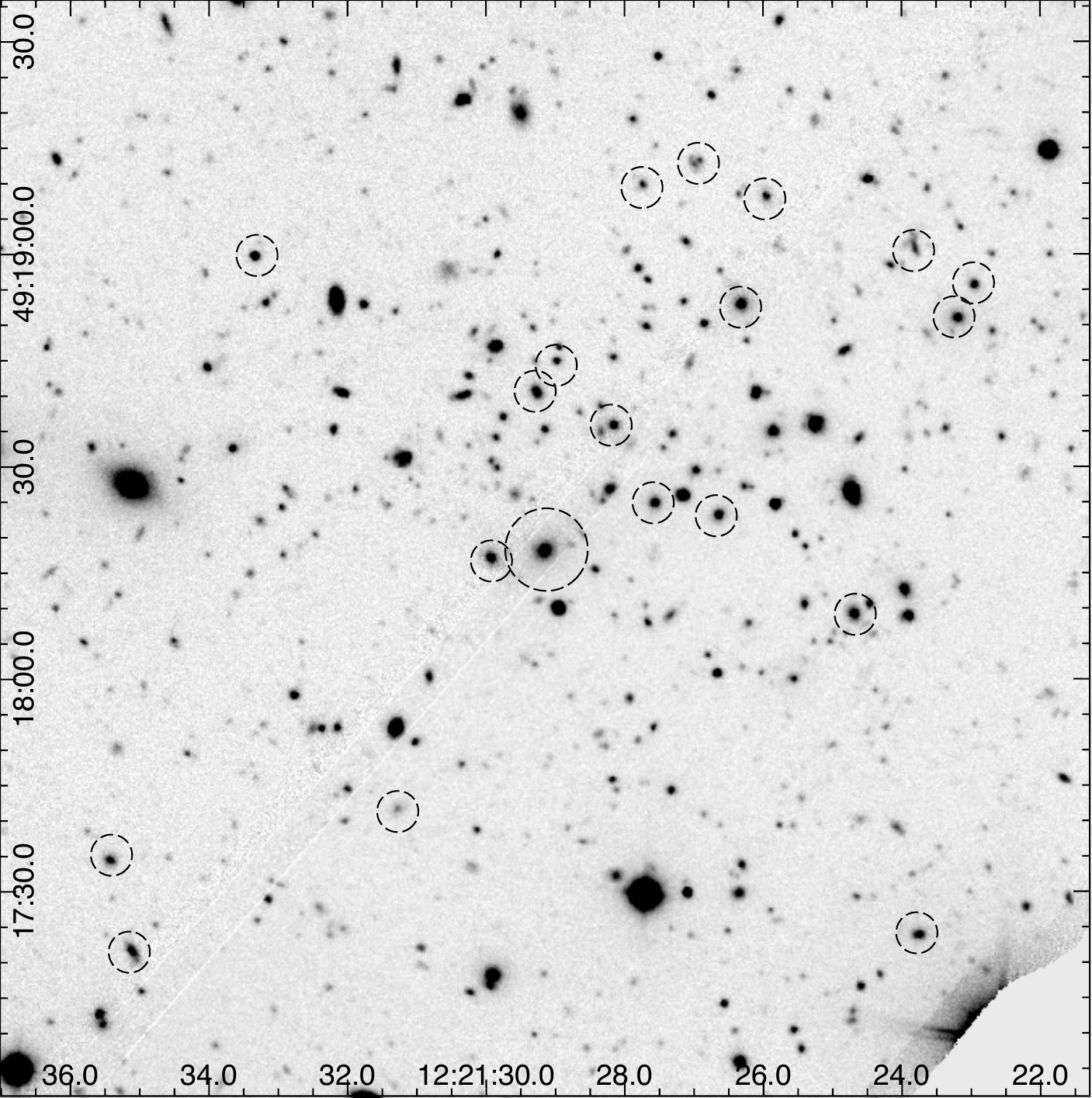}\\
\includegraphics[width=0.2\hsize]{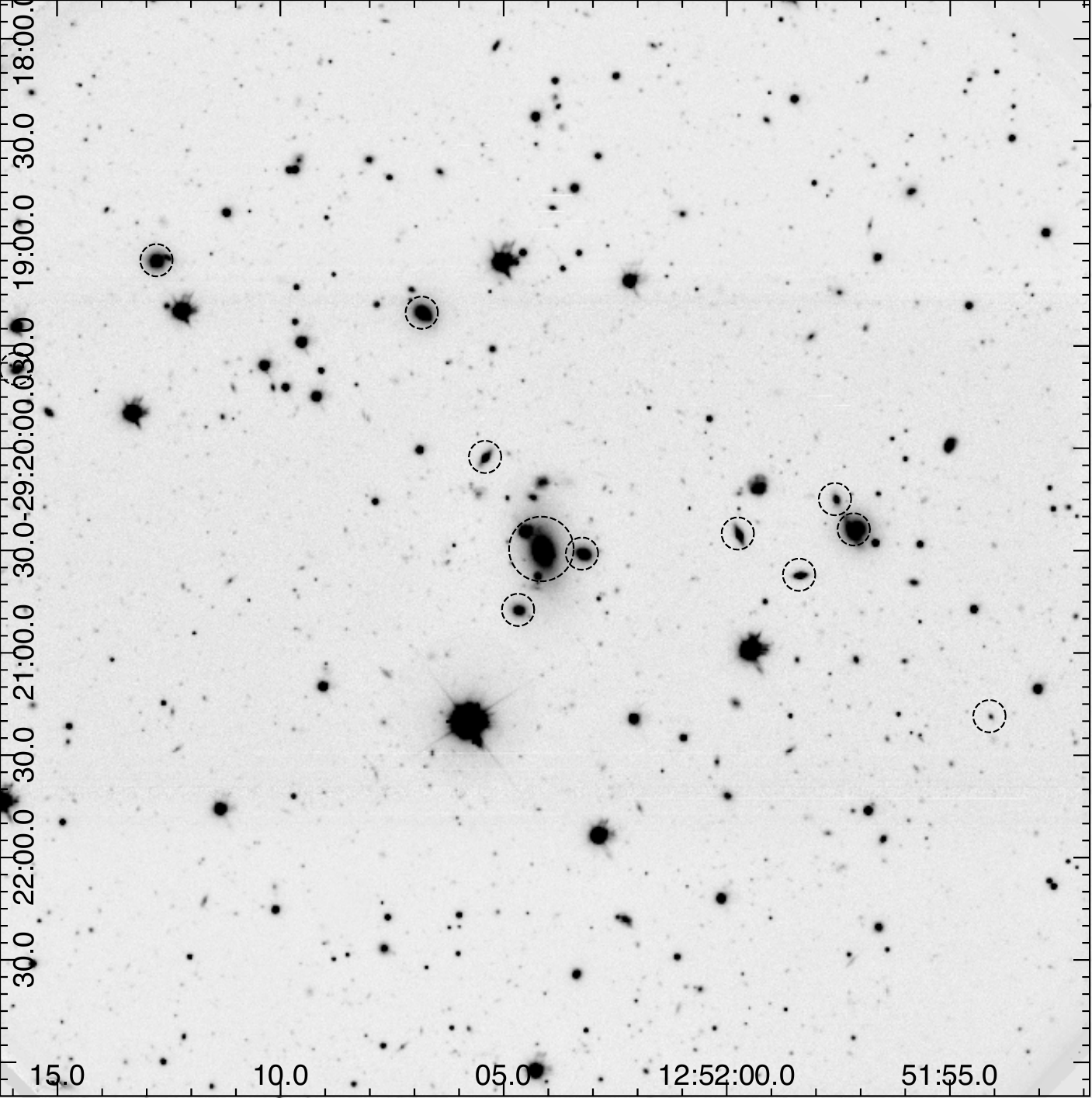}%
\includegraphics[width=0.2\hsize]{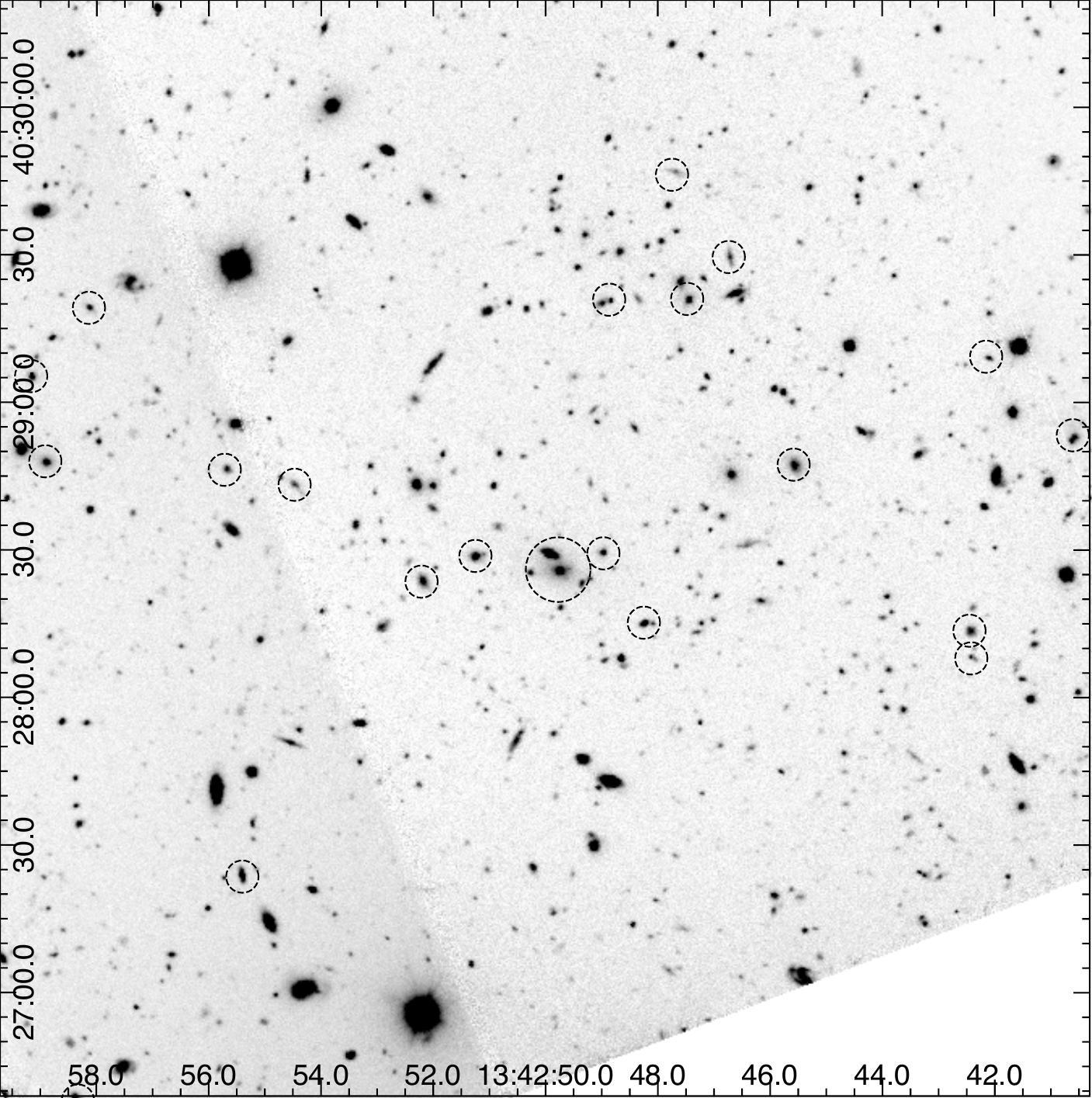}
	\caption{$r^\prime$ images of the low X-ray galaxy clusters observed 
with 
Gemini North and South telescopes: [VMF98]001, [VMF98]022, [VMF98]093, 
[VMF98]097, [VMF98]102, 
[VMF98]119, [VMF98]124 and [VMF98]148 (from upper left to lower right panels).
The images are of 1.5 Mpc side except for the clusters: [VMF98]022, [VMF98]093
and [VMF98]124 with 1 Mpc side.   North is up and 
East is to the left. }
\label{fov3}
\end{figure*}

\subsection{Data Reduction}

All the observations were reduced using standard procedures in IRAF\footnote{
IRAF is distributed by the National Optical Astronomy Observatories,
which are operated by the Association of Universities for Research
in Astronomy, Inc., under cooperative agreement with the National
Science Foundation.} (Tody 1993) and 
specific packages, depending on instruments and telescopes. The images were 
overscanned and 
bias subtracted, trimmed and flat-fielded following the standard reduction 
algorithms. Individual images were put into a common position system and then 
combined to create final images. Figures~\ref{fov1}, \ref{fov2} and \ref{fov3} 
show the $R$ or $r^\prime$ images 
of the galaxy cluster sample obtained with the LCO, CTIO and Gemini telescopes,
respectively.  They are images of 1.5 Mpc side except for   
[VMF98]022, [VMF98]093 and [VMF98]124 with a 1 Mpc side due to observational
constraints.    We have marked the
galaxy members (as addressed in Section~\ref{members}) and the BCG with 
circles.

\subsubsection{Source Detection and Photometry}
\label{photom}

Extracting faint objects from deep images was a major concern
in our study. The combination of SExtractor v2.19.5 (Bertin \& Arnouts, 1996) 
and PSFEx v3.17.1 (PSF Extractor, Bertin 2011) was used
with different configurations in order to detect sources
and to obtain the astrometric and photometric parameters, including 
position, magnitudes, colors and structural properties.  
SExtractor creates photometric catalogs from the observed images and PSFEx 
extracts models of the Point Spread Function (PSF) from the images 
processed by SExtractor. The generated PSF models are
used for model-fitting photometry and morphological analyses.  In general, 
SExtractor was run on  
the $r^\prime$ or $R$ passband images as reference, 
applying 
different Gaussian convolution filters, which depends on the image quality.
For bright 
detections in crowded central regions, filter width of 1.5 pix 
in 3$\times$3 pixels was used while for extended low-surface brightness 
objects in more
external parts, 2.0 pix in 5$\times$5 pixels was utilized. 
The minimum area for detections was defined with 7 pixels at lower redshifts 
and 4 pixels at higher redshifts.  We have considered as detected sources those
with 2 $\sigma$ above the detection limit. 
Deblending was performed with 16 sub-thresholds and a minimum contrast of 
0.005 in flux.  After running SExtractor, we have checked the detections 
aiming to
find spurious objects and false detections.  These are typically located in 
the outer regions of the CCDs and they were removed by hand. SExtractor 
was then
run in dual-image mode using the reference as the detection 
image.  With this methodology, objects in all filters have the 
same aperture size, which is appropriate for measuring colors.

\begin{figure}[!ht]
\begin{centering}
\includegraphics[width=90mm,height=90mm]{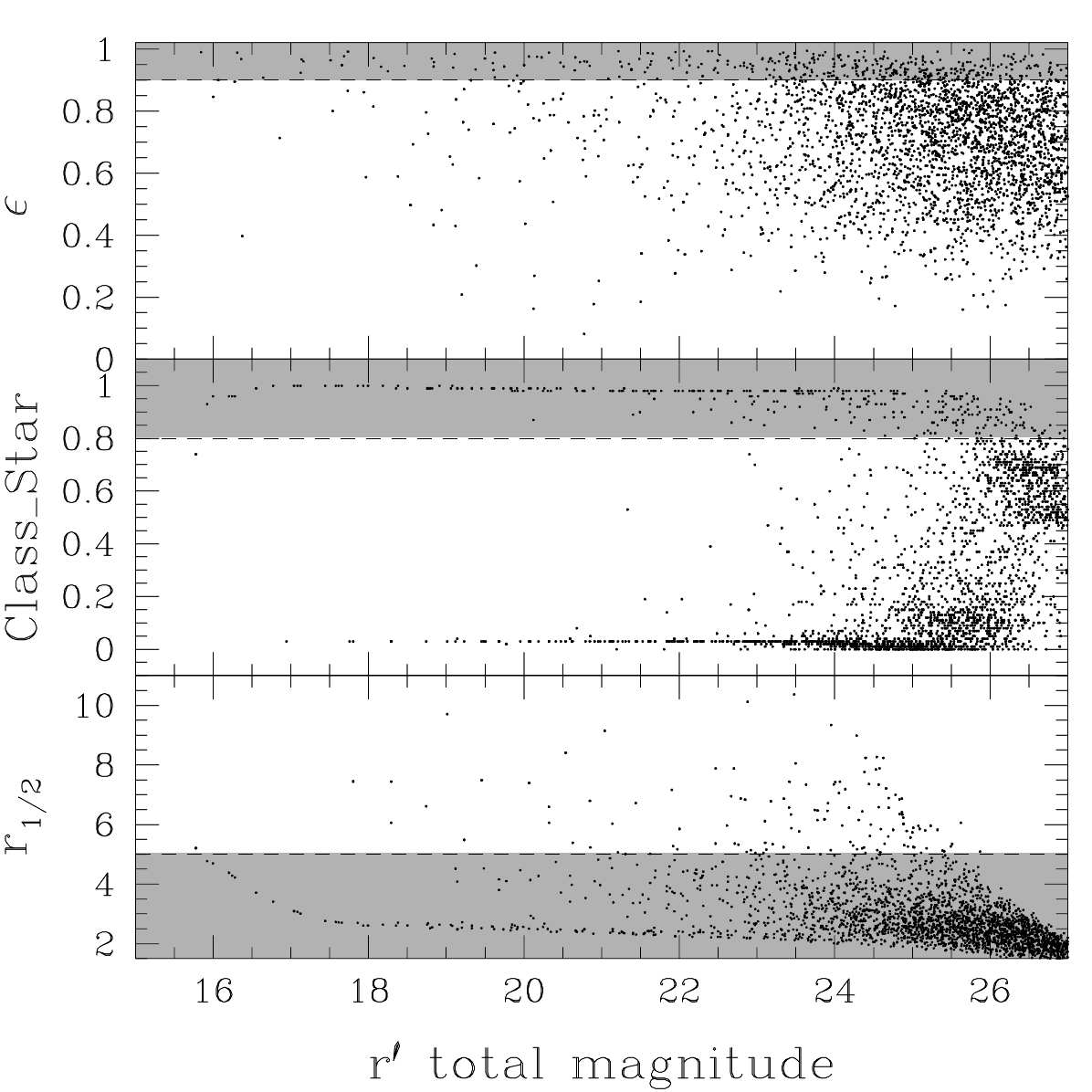}
\caption{The three adopted star-galaxy indicators versus $r^\prime$ total magnitudes. The gray zones indicate 
the typical regions where point sources are located. The dashed line marks our limit to separate galaxies and 
stars.} 
\label{stargalaxy}
\end{centering}
\end{figure}

The objects were classified by performing the star-galaxy 
separation using three different parameters: ellipticity 
($.pdfilon = 1 - b/a$); 
CLASS\_STAR and  half-light radius (r$_{1/2}$). $b/a$ is the axial ratio and
CLASS\_STAR is the 
SExtractor parameter 
associated with the light distribution of the detected objects.  Galaxies are
defined as those 
objects satisfying simultaneously $.pdfilon$ $<$ 0.9;  CLASS\_STAR $<$ 0.8 
and r$_{1/2}$ $>$ 5 pixels.  Figure~\ref{stargalaxy} shows an example of these 
parameters used to
define the galaxies in the cluster [VMF98]124: $.pdfilon$, 
CLASS\_STAR and r$_{1/2}$ in pixels as a function of 
$r^\prime$ total magnitudes.  The adopted criteria allow us to remove spurious
objects in the galaxy cluster fields, where saturated or overlapped objects
in projection were among the most frequent problems.

We have adopted PSF magnitudes as the galaxy total magnitude and aperture magnitudes to obtain colors. Using the same aperture 
diameter for the whole 
galaxy 
cluster sample may introduce some systematic effects  due to considering 
different
parts of the galaxies.
We have used an aperture size equivalent to 
a diameter of 10 kpc at the cluster redshift for aperture magnitudes.  This 
diameter is a compromise value that takes into account 
the typical galaxy size avoiding external contaminations.

\subsubsection{Magnitude calibration}

In order to check the photometric calibration, objects 
classified as stars obtained with SExtractor in the observed images 
were compared with the USNO-A2.0 catalogue (Monet et al. 2003) for B, R and I 
magnitudes; the  NOMAD catalogue (Zacharias et al. 2005) for V magnitudes
and the Sloan Digital 
Sky Survey - DR12 (Alam et al. 2015, hereafter SDSS) for $g^\prime$, $r^\prime$ 
and $i^\prime$ magnitudes.  In general, saturated stars or objects fainter 
than the catalog limiting magnitude were
not used as possible misclassifications may contribute with inaccurate 
magnitudes, particularly at fainter levels.  The galaxy cluster [VMF98]124 is not covered 
by the
 SDSS and we have first converted the USNO stellar 
magnitudes into the SDSS system using Fukugita et al. (1996) relations. 
Table~\ref{table2.2} shows these magnitude offsets corresponding to the 
different observed passbands for each galaxy cluster.
These values  were taken into account for the final magnitudes 
and colors.  The magnitudes are in the AB system and have been corrected for 
galactic extinction by using reddening maps from Schlegel et al. (1998) and the
Cardelli, Clayton \& Mathis (1989) relations.  

\begin{deluxetable*}{crrrrccr}
\tabletypesize{\scriptsize}
\tablewidth{0pc}
\tablecolumns{8}
\tablecaption{Photometric calibration: magnitude offsets.\label{table2.2}}
\tablehead{
\colhead{[VMF098]} & \colhead{$\Delta B$} & \colhead{$\Delta V$} & \colhead{$\Delta R$} & \colhead{$\Delta I$} & \colhead{$\Delta g^{\prime}$} & \colhead{$\Delta r^{\prime}$} & \colhead{$\Delta i^{\prime}$}\\
\colhead{Id.} & \colhead{} & \colhead{} & \colhead{} & \colhead{} & \colhead{} &	\colhead{} & \colhead{}}
\startdata
001 & -- & -- & -- & -- & -- & -0.090$\pm$0.059 & 0.137$\pm$0.042 \\
004 & 0.198$\pm$0.014 & 0.215$\pm$0.096 & 0.012$\pm$0.069 & -0.274$\pm$0.052 & -- & -- & -- \\
011 & 0.040$\pm$0.007 & -- & -0.102$\pm$0.024 & -- & -- & -- & -- \\
015 & -0.113$\pm$0.010 & 0.101$\pm$0.032 & -0.102$\pm$0.026 & -0.224$\pm$0.023 & -- & -- & -- \\
022 & -- & -- & -- & -- & -- & 0.276$\pm$0.019 & -0.244$\pm$0.027 \\
024 & 0.178$\pm$0.004 & 0.193$\pm$0.018 & -0.111$\pm$0.038 & 0.268$\pm$0.039 & -- & -- & -- \\
025 & 0.212$\pm$0.011 & 0.177$\pm$0.046 & -0.285$\pm$0.033 & -0.099$\pm$0.006 & -- & -- & -- \\
045 & 0.030$\pm$0.091 & -0.265$\pm$0.024 & 0.142$\pm$0.025 & 0.160$\pm$0.037 & -- & -- & -- \\
046 & 0.101$\pm$0.211 & 0.244$\pm$0.217 & -0.137$\pm$0.021 & -0.212$\pm$0.050 & -- & -- & -- \\
093 & -- & -- & -- & -- & -- & 0.181$\pm$0.014 & 0.264$\pm$0.016 \\
097 & -- & -- & -- & -- & 0.221$\pm$0.089 & 0.299$\pm$0.061 & -- \\
102 & -- & -- & -- & -- & -- & 0.114$\pm$0.091 & -- \\
113 & 0.052$\pm$0.511 & 0.091$\pm$0.055 & 0.103$\pm$0.018 & 0.099$\pm$0.027 & -- & -- & -- \\
119 & -- & -- & -- & -- & -- & 0.049$\pm$0.084 & 0.040$\pm$0.101 \\
124 & -- & -- & -- & -- & 0.222$\pm$0.027 & 0.060$\pm$0.011 & -- \\
148 & -- & -- & -- & -- & -- & 0.323$\pm$0.049 & 0.224$\pm$0.170 \\
211 & 0.026$\pm$0.022 & -- & 0.278$\pm$0.042 & -- & -- & -- & -- \\
214 & 0.300$\pm$0.290 & 0.193$\pm$0.015 & -0.111$\pm$0.040 & 0.268$\pm$0.039 & -- & -- & -- \\
215 & 0.211$\pm$0.167 & 0.108$\pm$0.048 & -0.123$\pm$0.006 & 0.221$\pm$0.009 & -- & -- & -- \\
\enddata
\end{deluxetable*}

\subsubsection{Limiting Magnitudes and Completeness}
\label{limitingmag}

In order to check the SExtractor behavior at fainter magnitudes, we
have estimated magnitude limits and completeness levels using 
simulated catalogues and images created with the 
Astromatic packages STUFF and 
SKYMAKER (Bertin 2009). STUFF simulates field galaxy catalogues in a Poisson 
distribution, for different redshift slices from 0 $<$ z $<$ 20, with the 
number of
galaxies and their absolute luminosities being taken from a non-evolving 
Schechter Luminosity Function (Schechter 1976). 
Galaxy profiles were modelled 
by the contribution of two components: a de Vaucouleurs bulge and an exponential
disk (for details, see Erben et al. 2001).  The photometric,
structural and astrometric parameters for all objects were generated in all
passbands using the filter transmission 
curves  and 
spectral energy distributions.  
SKYMAKER produces
realistic ground-based Point Spread Functions using STUFF catalogs taking 
into account the
instrumentation and observing conditions. 

\begin{figure}[!ht]
\begin{centering}
\includegraphics[width=90mm,height=90mm]{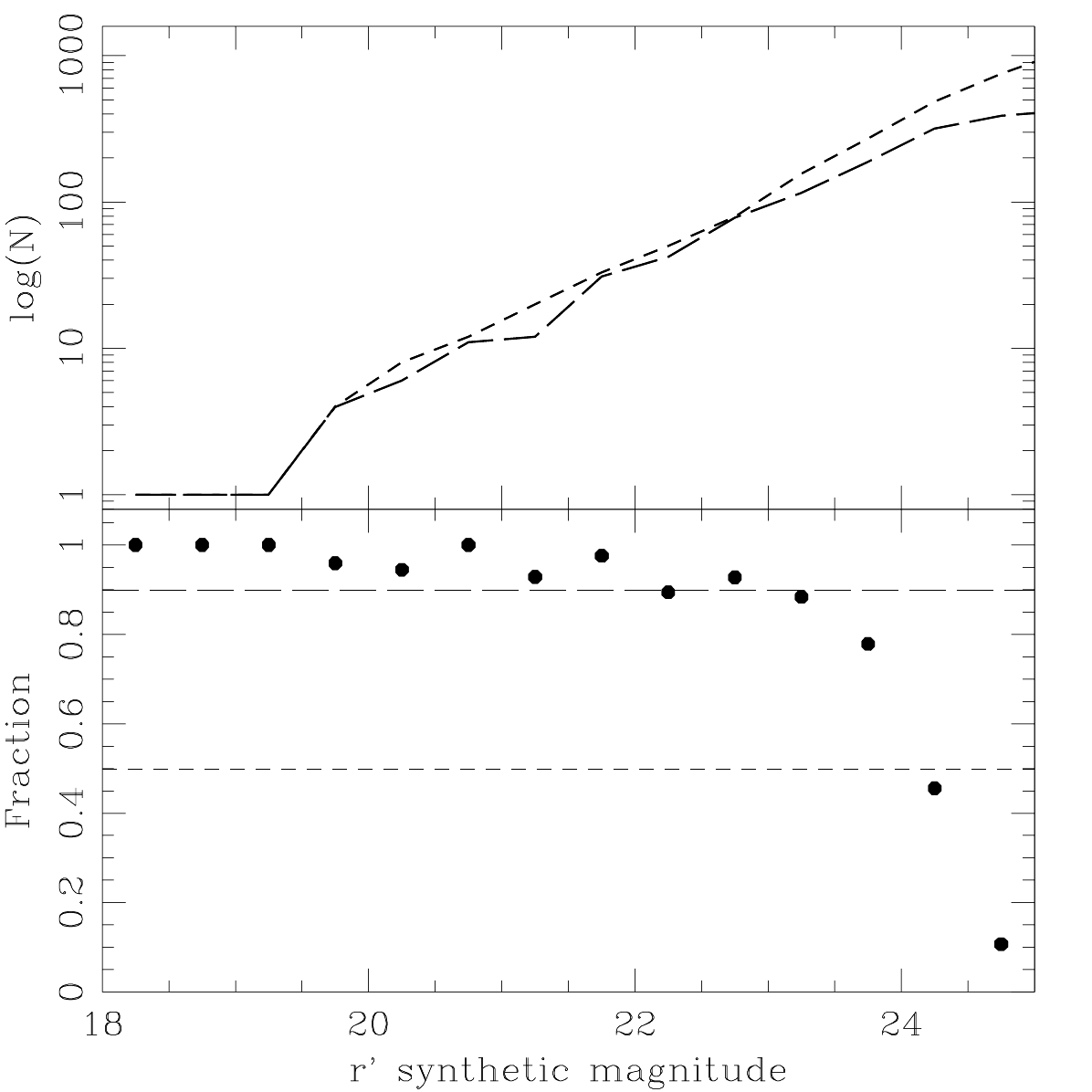}
\caption{Completeness levels using simulated data.  The upper panel shows
the SKYMAKER (short dash
line) and SExtractor synthetic (long dash line) total magnitude 
distributions in logarithmic scale.  In the lower panel, the ratio of 
these two distributions are shown. The two horizontal lines correspond to 50\% 
and 90\%  completeness
levels.} 
\label{completeness}
\end{centering}
\end{figure}

Using these packages, we have reproduced our observations generating 
synthetic images and catalogues. We have run 
SExtractor in these
simulated images and the resulting catalogues were compared with
those created by SKYMAKER. 
Figure~\ref{completeness} shows an example of the number of object detections
per magnitude bin (upper panel) obtained from the $r^{\prime}$ simulated 
catalogue and 
those sources detected by SExtractor from the synthetic images.  The 
distributions are shown in logarithmic scale as short and long dash lines, 
respectively.   In the lower panel, the fraction of these detected 
distributions per magnitude bin are displayed indicating the 50\% and 
 90\% completeness fractions.  This figure indicates that the number of 
detections are similar; and in this example, SKYMAKER and 
SExtractor magnitudes are in good mutual agreement up to $r^{\prime}$ of 
about 23 mag.  For each galaxy cluster, the observing conditions were 
simulated with this procedure and 
magnitude limits were obtained.  Table~\ref{table2.3} shows the limiting 
magnitudes within 90\% completeness. 

As the magnitude errors provided by SExtractor are underestimated 
(White et al. 2005), our error estimates were based on the comparison between the
synthetic magnitudes derived from SKYMAKER and those obtained with SExtractor.
Within a 90\% completeness level, the mean magnitude errors were found to be
approximately 0.1 mag.

\begin{deluxetable}{crrrrccr}
\tabletypesize{\scriptsize}
\tablewidth{0pc}
\tablecolumns{8}
\tablecaption{Observed magnitude limits within 90\% completeness levels.\label{table2.3}}
\tablehead{
\colhead{[VMF098]} & \colhead{$B$} & \colhead{$V$} & \colhead{$R$} & \colhead{$I$} & \colhead{$g^{\prime}$} & \colhead{$r^{\prime}$} & \colhead{$i^{\prime}$}\\
\colhead{Id.} & \colhead{} & \colhead{} & \colhead{} & \colhead{} & \colhead{} &	\colhead{} & \colhead{}}
\startdata
001 & -- & -- & -- & -- & -- & 23.54 & 23.21 \\
004 & 23.90 & 22.89 & 22.86 & 20.90 & -- & -- & -- \\
011 & 23.87 & -- & 22.90 & -- & -- & -- & -- \\
015 & 23.77 & 22.93 & 22.40 & 21.41 & -- & -- & -- \\
022 & -- & -- & -- & -- & -- & 23.16 & 23.03 \\
024 & 23.39 & 22.86 & 22.09 & 21.90 & -- & -- & -- \\
025 & 23.43 & 22.80 & 22.13 & 21.94 & -- & -- & -- \\
045 & 22.90 & 22.93 & 22.71 & 21.71 & -- & -- & -- \\
046 & 23.01 & 22.88 & 22.70 & 21.62 & -- & -- & -- \\
093 & -- & -- & -- & -- & -- & 23.62 & 23.51 \\
097 & -- & -- & -- & -- & 23.75	& 23.23	& -- \\
102 & -- & -- & -- & -- & -- & 23.01 & -- \\
113 & 22.96 & 22.72 & 22.55 & 21.87 & -- & -- & -- \\
119 & -- & -- & -- & -- & -- & 23.70 & 23.00  \\
124 & -- & -- & -- & -- & 23.09	& 23.05 & -- \\
148 & -- & -- & -- & -- & -- & 23.34 & 22.93 \\
211 & 23.11 & -- & 22.14 & -- & -- & -- & -- \\
214 & 23.66 & 22.47 & 22.16 & 20.67 & -- & -- & -- \\
215 & 23.39 & 22.43 & 21.90 & 20.94 & -- & -- & -- \\
\enddata
\end{deluxetable}

\section{Spectroscopy}

\subsection{Observations}

The GMOS instrument was used 
in the MOS mode at the Gemini North and South telescopes 
during the SVP in 2003, under photometric conditions with 
typical seeing of about 0.6 and 0.9 arcsec.
A GMOS grating of 400 lines/mm ruling density centered at 
6700 \AA\ was used, covering a wavelength range of 4400 to 9800 \AA. 
The spectra had a resolution of about 5.5 \AA, with a dispersion of 
1.37 \AA/pix, and offsets of $\sim$ 35 \AA\ were applied between 
exposures in the 
spectral direction toward the blue and/or the red to fill the gaps between 
CCDs. The comparison lamp (CuAR) spectra 
were taken after each science exposure. 

We have obtained spectroscopic data 
for galaxies in the fields of 
four galaxy clusters: [VMF98]022, 
[VMF98]097, [VMF98]102 and [VMF98]124.  The spectroscopic targets were based 
on the photometric 
catalogues generated with SExtactor, described in the
previous section.   
In order to study cluster galaxy population, we selected objects brighter than
$r^\prime = $23 mag, without any color criteria (Carrasco et al. 
2007).  
Observations were performed with two masks for the clusters [VMF98]097 
and [VMF98]102. Objects brighter than $r^\prime$ = 20 mag were observed in a
single mask with shorter exposure time than the fainter ones.  
Table~\ref{table3} shows the observed cluster 
identification, with a summary of the spectroscopic observations 
including observation date, exposure time and number of observed spectra per 
mask.

\begin{deluxetable*}{ccccccccc}
\tabletypesize{\scriptsize}
\tablewidth{0pc}
\tablecolumns{9}
\tablecaption{Spectroscopic observations and results.\label{table3}}
\tablehead{
\colhead{[VMF98]} & \colhead{Observation}  &  \colhead{Exposure}   & \colhead{Observed}    &   \colhead{Member} & \colhead{Mean} & \colhead{Velocity}  &   \colhead{Mean} & \colhead{Velocity}  \\
\colhead{Id.} & \colhead{Date} & \colhead{Time} & \colhead{Spectra} & \colhead{galaxies} & \colhead{redshift\tablenotemark{1}} & \colhead{dispersion\tablenotemark{1}} & \colhead{redshift\tablenotemark{2}} & \colhead{dispersion\tablenotemark{2}}} 
\startdata
022  	& 09/26/2003	  &   2400      & 51 & 26 & 0.247 & 508 & 0.248 & 412$\pm$81\\
097  	& 05/28/2003	  &   3600      & 22 & 37 & 0.482 & 775 & 0.486 & 1970$\pm$328 \\
       	& 05/29/2003	  &   6000      & 53 &    &   &        \\ 
102  	& 05/26/2003	  &   3600      & 37 & 22 & 0.409 & 675 & 0.410 & 675$\pm$169 \\
       	& 05/31/2003	  &   6000      & 37 & &      &        \\
124  	& 05/24/2003	  &   1800      & 29 & 12 & 0.185 & 700 & 0.185 & 681$\pm$197\\ 
\enddata
\tablenotetext{1}{Value obtained using the bi-weight $\sigma$ estimates.}
\tablenotetext{2}{Value obtained using the the jackknife error estimates.}
\end{deluxetable*}

\subsection{Data Reduction}

The observations including comparison lamps and spectroscopic flats were bias subtracted and 
trimmed using the Gemini IRAF package.  The flats were processed by removing the calibration unit plus the GMOS spectral response and the calibration 
unit uneven illumination, which were then normalized to leave only the 
pixel-to-pixel variation and fringing.  Details of the reduction procedure are found in Carrasco et al. (2007). 

The procedure to measure the galaxy radial velocity 
was started with an inspection of the galaxy 
spectra searching for strong features such as absorption and/or emission 
lines. RVIDLINES was applied for galaxies with clear 
emission lines, identifying one or more spectral 
lines and comparing with the rest-frame
 wavelengths.  The average wavelength shifts were computed and 
converted to a radial velocity, with the residual of all 
shifts being used to estimate errors.  In contrast,
FXCOR was 
applied in early-type galaxies, cross-correlating the observed spectra with 
high signal-to-noise 
templates, with the R-value (Tonry \& Davis, 1979) used to define the quality 
of the 
measured radial velocities (Carrasco et al. 2007). 
For $R > 3.5$, the observed radial velocity was associated to the template that 
produced lower uncertainties.  However, in the case of R $\leq$ 3.5, absorption 
features were
searched for, and line-by-line Gaussian fits were obtained. Both RVIDLINES and 
FXCOR routines are part of the RV package in IRAF.  

We have obtained radial velocities of objects selected in the neighborhoods of the four galaxy
clusters and, for further analysis we need to know the completeness levels of
their magnitude distributions.  For the photometric samples, the limiting magnitudes and 
completeness levels are extensively discussed in section 
$\S$\ref{limitingmag}.  For the spectroscopy, the magnitude distributions 
 of the clusters [VMF98]022, [VMF98]097 and [VMF98]102
show a brighter limiting magnitude of  $r^\prime \sim$  20.5 mag, reaching a 
90\% completeness levels similar to the 
photometric samples. The cluster [VMF98]124 has been observed with only one mask
with shorter exposure time, resulting in about 50\% completeness at the same
limiting magnitude. 

\section{Cluster membership}\label{members}

In order to understand better the cluster assembly and
morphological evolution of galaxies in low X-ray luminosity clusters,
it is crucial to define cluster membership.  Even when precise radial 
velocities are available for a large number of objects, the galaxy assignment 
of a 
cluster is not guaranteed. In effect, relatively distant infalling galaxies 
onto the cluster will 
appear closer to the cluster kinematic center. On the other hand, interlopers, 
namely galaxies that are not confined to the cluster region may appear in 
projection with a low clustercentric relative velocity and therefore could be 
wrongly assigned as members.
 Using mock galaxy redshift surveys, van den 
Bosch et al. (2004) found that the velocity distribution 
of interlopers is strongly peaked towards the cluster mean radial 
velocity, thus introducing a bias 
in the cluster member assignment, especially at higher velocity dispersions.

We have selected galaxies within a projected radius of
0.75 Mpc from the X-ray emission peak.  This choice is based on  
 the angular size of the lowest redshift clusters and the available instrument FOVs.
 The relatively small radius minimizes 
foreground/background galaxy contamination and correspond to the densest
cluster regions.  It must be noted that by exploring the central regions
of galaxy clusters, our study will focus on those galaxies mostly affected
by the cluster environment at these redshifts.  Thus, our analysis does not
address properties of galaxies far from the cluster core which could be
potentially different than those in the higher density regions.

\subsection{Spectroscopic membership and substructures}

In our cluster sample, only 
four galaxy clusters: [VMF98]022, [VMF98]097, [VMF98]102 
and [VMF98]124 have spectroscopic redshift measurements. 
The  
cluster membership was defined using the projected radius and also the 
spectroscopic restriction  
$\Delta$V $<  \sigma$,
with $\Delta$V defined as the difference between a given 
galaxy radial velocity 
 and the cluster  redshift given by Mullis et al. (2003).
Testing membership 
assignment
using 1 or 2 $\sigma$s show in general, small
differences between the samples.  However, there are some variations in the 
cluster [VMF98]097 with  signs of a more 
complex morphology (Carrasco et al. 2007, Nilo Castell\'on et al. 2014).
As previously mentioned, although membership cannot
be totally guaranteed, we believe that the use of 1 $\sigma$ restriction 
is appropriate to minimize interlopers.  The number of cluster members
are shown in Column 5 of the Table~\ref{table3}.

The bi-weight estimator (Beers, Flynn \& Gebhardt 1990) is 
statistically more robust and 
efficient for computing the central location of the redshift distribution 
than the standard mean. 
Biviano et al. (2006) have used this
estimator for galaxy clusters with more than 15 members.  
We have obtained bi-weight $\sigma$ estimates for the galaxy clusters 
with spectroscopic measurements. Using cluster members, columns 6 and 7 of 
Table~\ref{table3} shows  the mean redshift and the 
line-of-sight velocity 
dispersion obtained with this estimate.  The uncertainties  
derived from a bootstrap resampling 
technique were approximately 0.001 for redshifts and 90 km~s$^{-1}$ for
velocity dispersions.   We have also computed with cluster members, 
the mean redshift and 
velocity dispersion using the jackknife error estimates, 
which are shown in columns 8 and 9, respectively.
We can see from this table that the velocity dispersion values using
the two estimators are in agreement within the uncertainties, except for
the galaxy cluster  [VMF98]097.  The higher value obtained
using the jackknife estimate is a consequence of the complex cluster 
morphology.   In the cluster [VMF98]124 which has only 12 
galaxy
members, we have also used
the "gapper" statistics 
obtaining a higher $\sigma \sim$  751 km~s$^{-1}$, which is in agreement
with the other estimates. 

\begin{figure}[!ht]
\begin{centering}
\includegraphics[width=0.48\hsize]{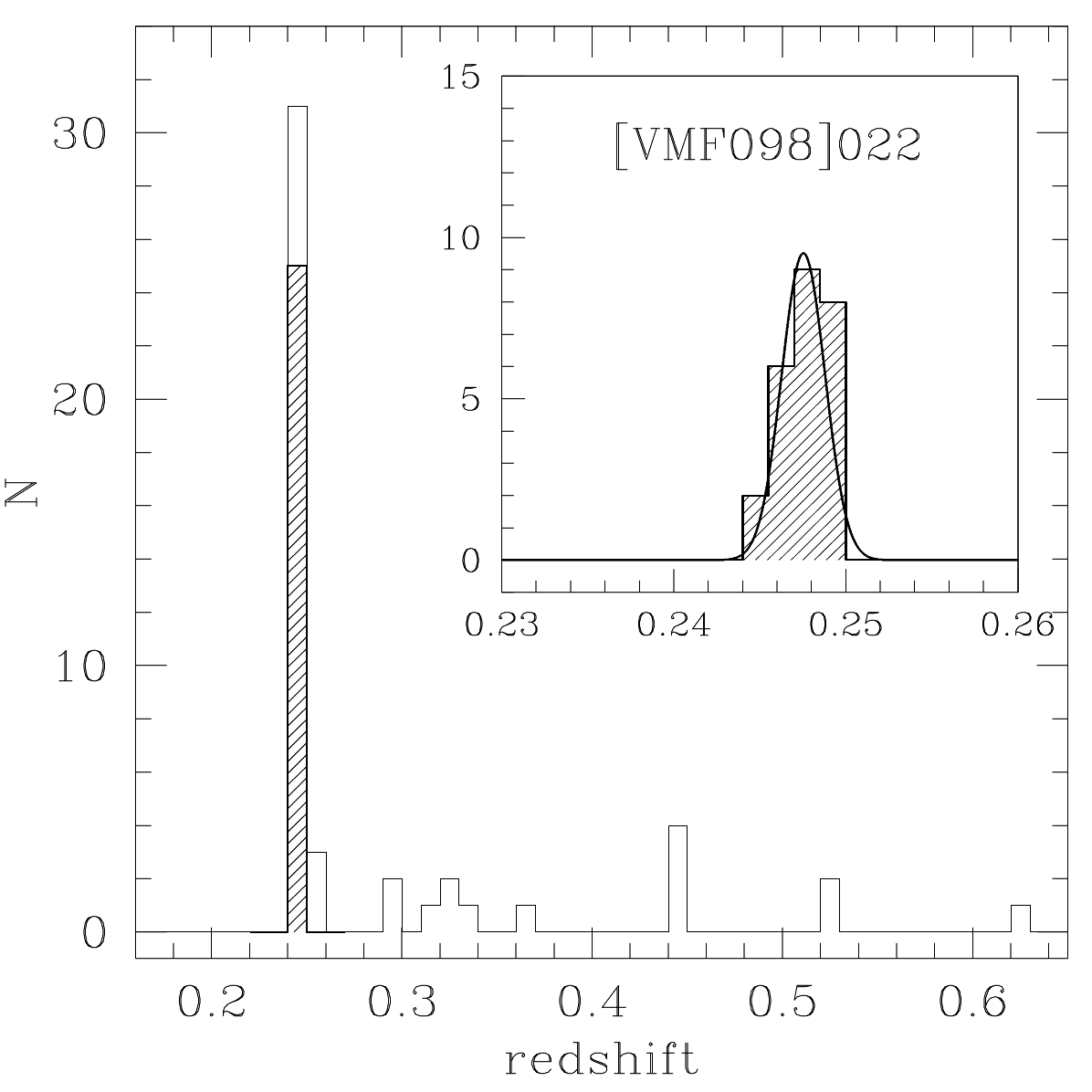}%
\includegraphics[width=0.48\hsize]{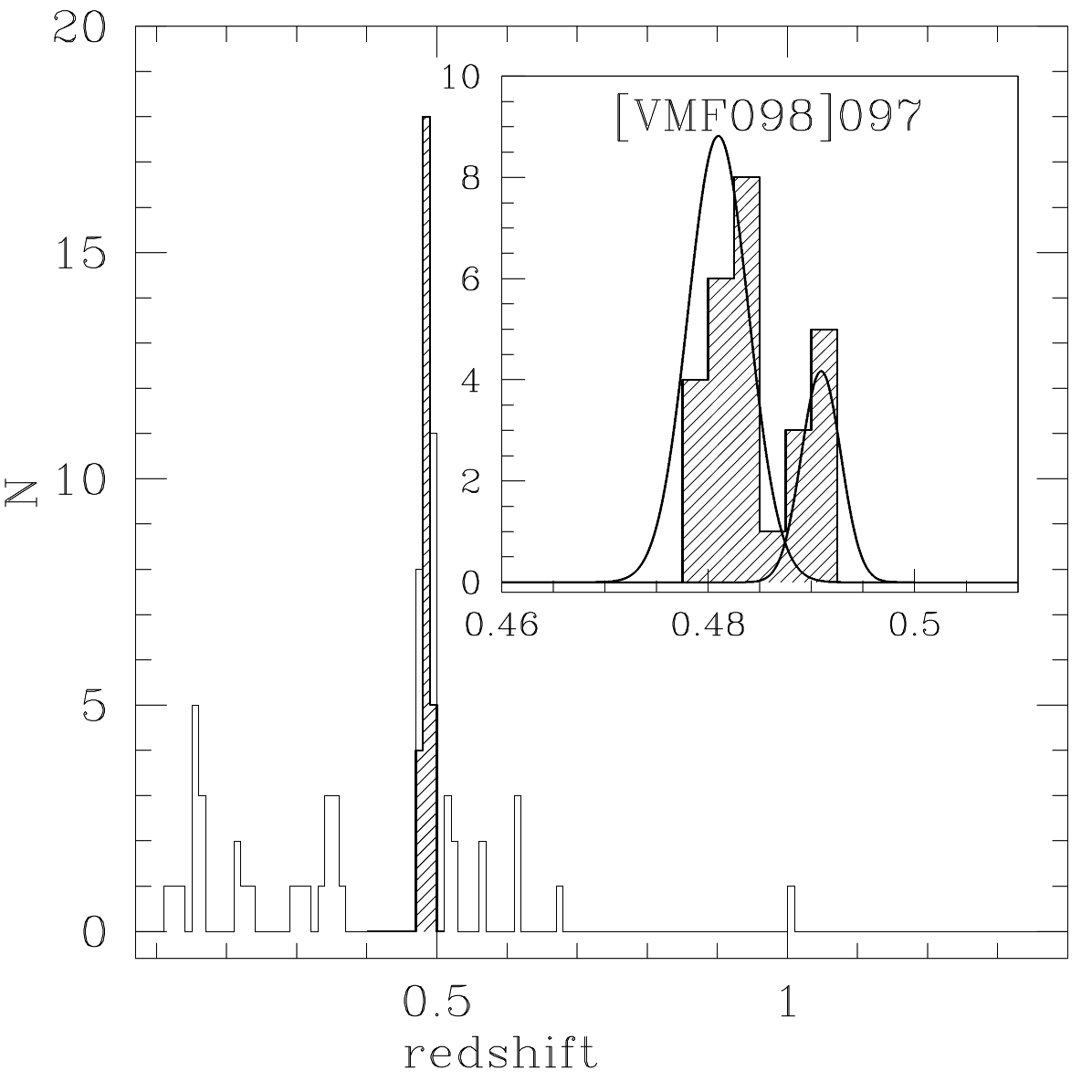}\\
\includegraphics[width=0.48\hsize]{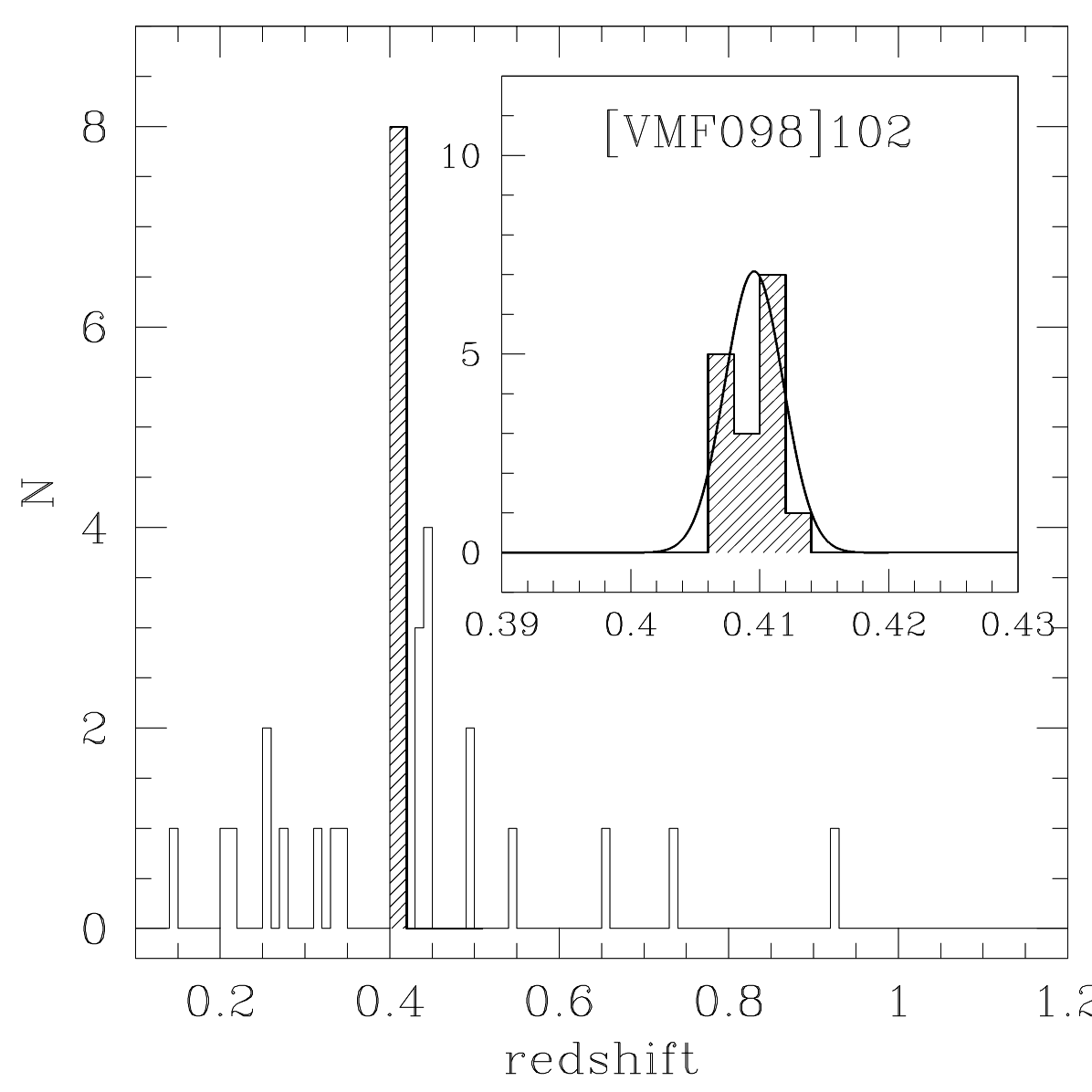}%
\includegraphics[width=0.48\hsize]{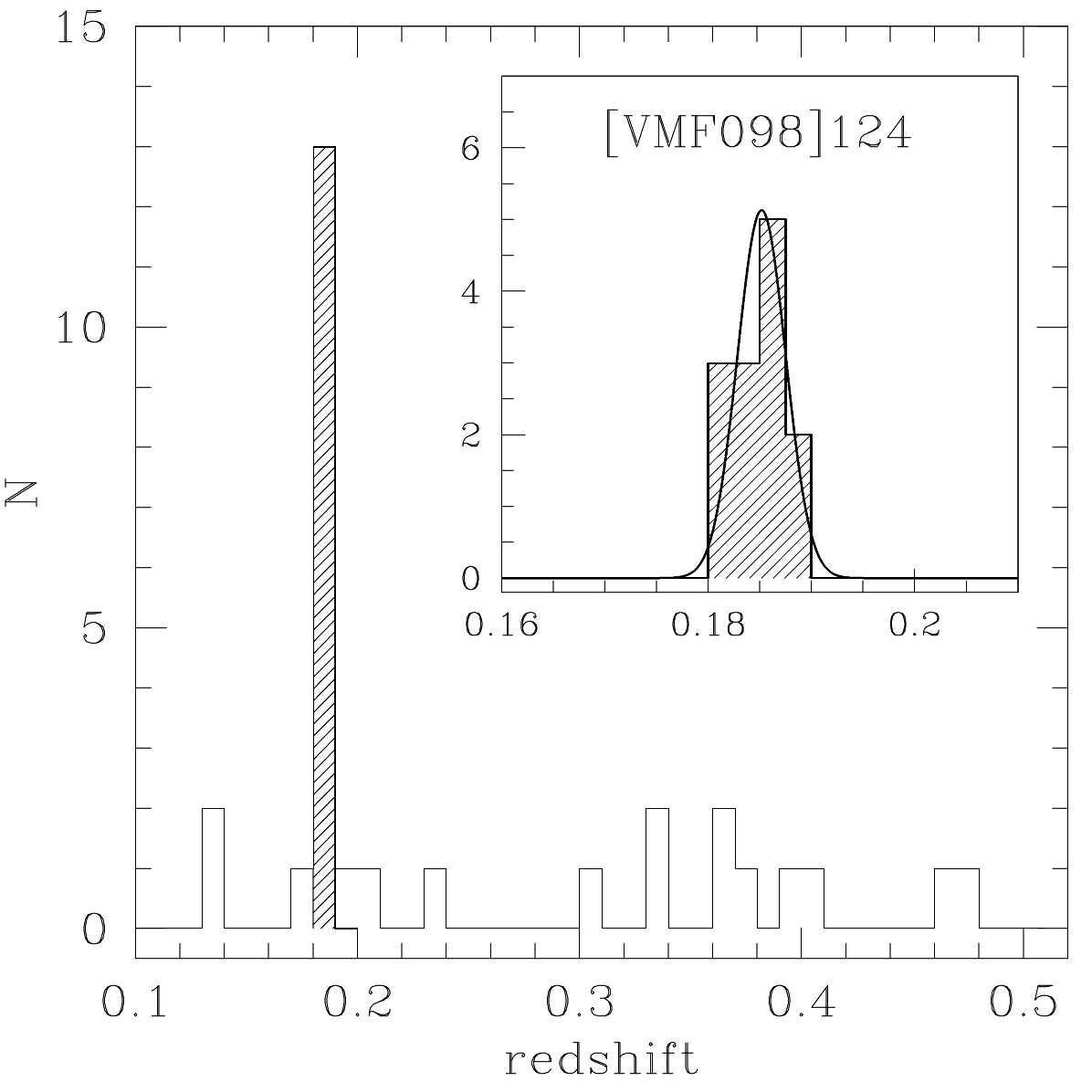}
\caption{Redshift distribution in the fields  of the four galaxy clusters
with available spectroscopy.   
Shaded histograms correspond to the distribution within 1$\sigma$. Right
corner panels show a detail of these distributions 
around the cluster redshift and the Gaussian fits.}
\label{redshift}
\end{centering}
\end{figure}

Figure~\ref{redshift} shows the 
observed redshift distribution in the neighborhoods 
of the four galaxy clusters with available spectroscopy, where the shaded 
parts correspond to the distributions 
within 1$\sigma$ of the mean cluster redshift. Gaussian function fits 
provide a suitable approximation to the line-of-sight
distribution. 
The foreground and background structures are also present in the 
figure. In
the right corner, a detail of this distribution and the 
Gaussian fit are displayed.  In the case of [VMF98]097, two peaks 
at redshifts 0.482 and 0.494 are clearly 
observed. This is the only galaxy cluster 
with well defined substructure, as also reported by Carrasco et al. (2007).
For the cluster [VMF98]102, 
there is a second peak corresponding to a probable background
system in the line-of-sight.
On average,  the uncertainties of the radial velocities were less 
than 55 km~s$^{-1}$.  

\subsection{Photometric redshifts}

The determination of cluster members
using spectroscopic redshifts is certainly the most accurate method 
but it is highly more expensive in telescope time,
especially at
fainter magnitudes. 
Colors trace the spectral energy distribution of galaxies 
at 
different redshifts and the intersection of a given set of observed colors with the  
allowed redshift ranges can be used to assess the redshift of a galaxy.
For this reason, redshift estimates of large
and deep samples of galaxy clusters can be obtained by using broad band 
photometry.  

Photometric redshift estimates
have been widely
used as an efficient way to study the galaxy
properties using statistical tools (Koo 1985; Connolly et al.
1995; Gwyn \& Hartwick 1996; Hogg et al. 1998; Fern\'andez-Soto et al.
1999; Ben\'itez 2000; Csabai et al. 2000; Budav\'ari et al. 2000), as for
example luminosity, colors and morphology.
Even with larger  uncertainties, they provide a powerful tool for studying
evolutionary galaxy properties of faint galaxies.
Two groups of  methods are used to estimate photometric redshifts. 
The template fitting technique makes use
of a small set of model galaxy
spectra derived from the $\chi^2$-based spectral
template-fitting package (Ben\'itez 1998; Bolzonella et al. 2000; Csabai et
al. 2003).  This approach consists in the reconstruction
of the observed galaxy colors in order to
find the best combination of template spectra at different redshifts.
The main disadvantage of this method is the relatively small number of
available templates in the library for different passbands. 
The second group is the empirical fitting technique which is based on empirical data (Connolly et al. 1995; 
Brunner et al. 1999; Collister \& Lahav 2004) requiring a large amount of a priori redshift 
information (training set), which may be in some cases a disadvantage. 
However, the main goal
of this procedure is to obtain redshift estimates as a
function of the photometric parameters as inferred from the training set.  

We consider photometric redshift estimates (Photo-z) to assign  
membership for the fifteen galaxy clusters without 
spectroscopic measurements.

\subsubsection{Photo-z with ANNz}

We have used the Artificial Neural Network
(ANNz, Collister \& Lahav 2004), one of the empirical fitting method to 
obtain photo-z, as described in 
O'Mill et al. (2012).  To calibrate the code, we have used 12280 galaxies with
spectroscopic redshifts derived from the
Canadian Network for Observational Cosmology (Yee, Ellingson \&
Carlberg, 1996, CNOC). This dataset was
randomly divided into two subsamples, thereby generating the
training and validation set.  

The ANNz produces better estimates when more observations in different
passbands are available.  We have used aperture magnitudes as defined in 
section~\ref{photom} for the nine galaxy clusters
observed in four passbands: $B, V, R$ and $I$ at the LCO and CTIO telescopes,
defining a 
subsample of galaxies reaching the CNOC limiting magnitudes.  
The
photometric
resdhift estimates take into account the telescope characteristics
and the adopted filters 
through the training and validation sets. 
The resulting ANNz architecture adopted here was $4:8:8:1$. Therefore, they 
were obtained using the photometric catalogs and
the distributions are related to the limiting magnitudes and completeness levels of these photometric samples (section $\S$\ref{limitingmag}).
Abdalla et al. (2011) and Dahlen et al. (2013) have discussed
associated bias and related uncertainties in the photometric redshift
estimates obtained with different
methods by a comparison with spectroscopic measurements. 
Abdalla et al. (2011) have found that the ANNz method
has an almost constant, small bias and a 1$\sigma$ scatter of about 0.06.
  The clusters
[VMF98]011 and [VMF98]211 have been observed at LCO in only two passbands
($B$ and $R$) and the photometric redshifts using ANNz are not accurate enough
for our purposes of membership and they are not consider in this work.

\subsubsection{Photo-z from the SDSS}

\begin{figure}!ht]
\begin{centering}
\includegraphics[width=0.48\hsize]{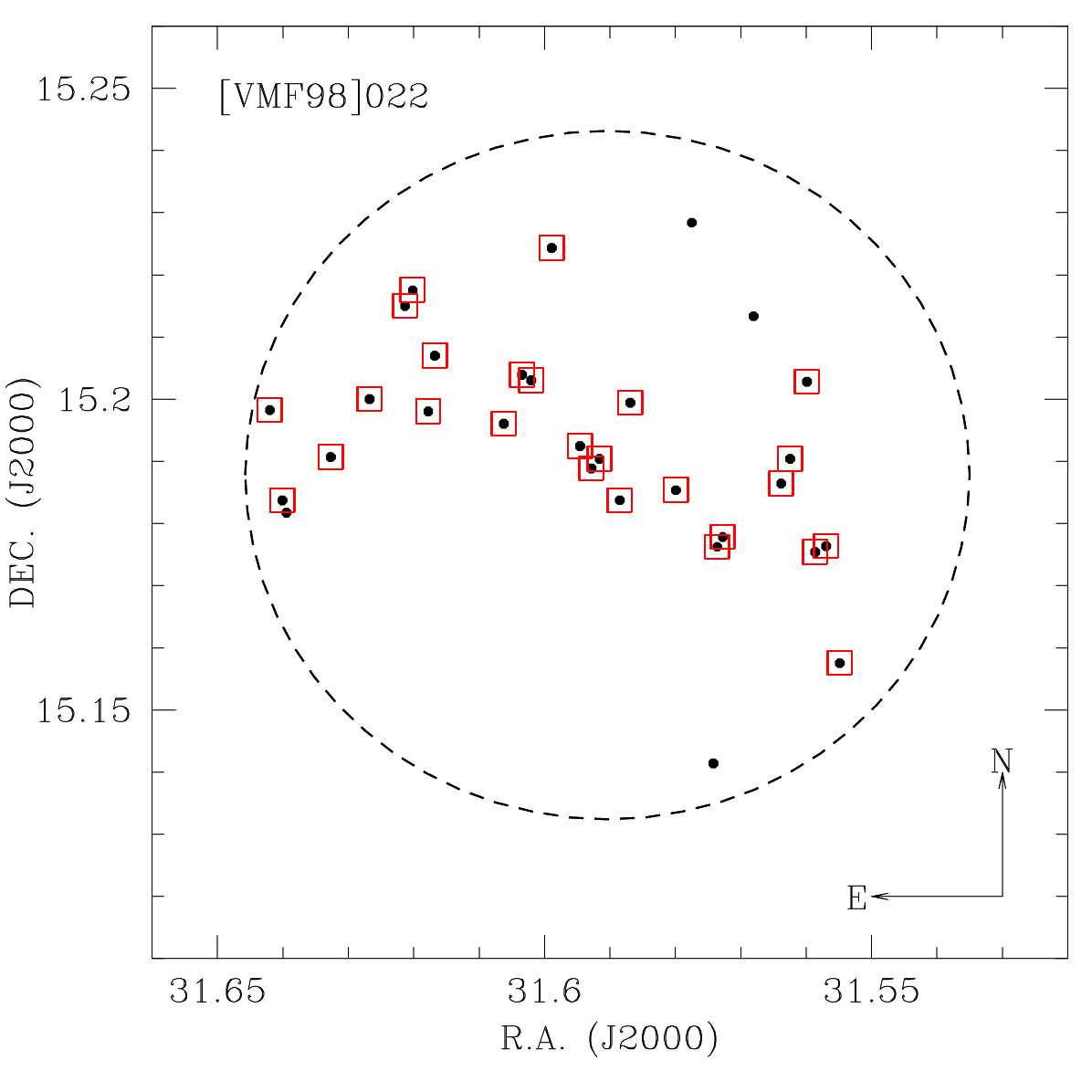}%
\includegraphics[width=0.48\hsize]{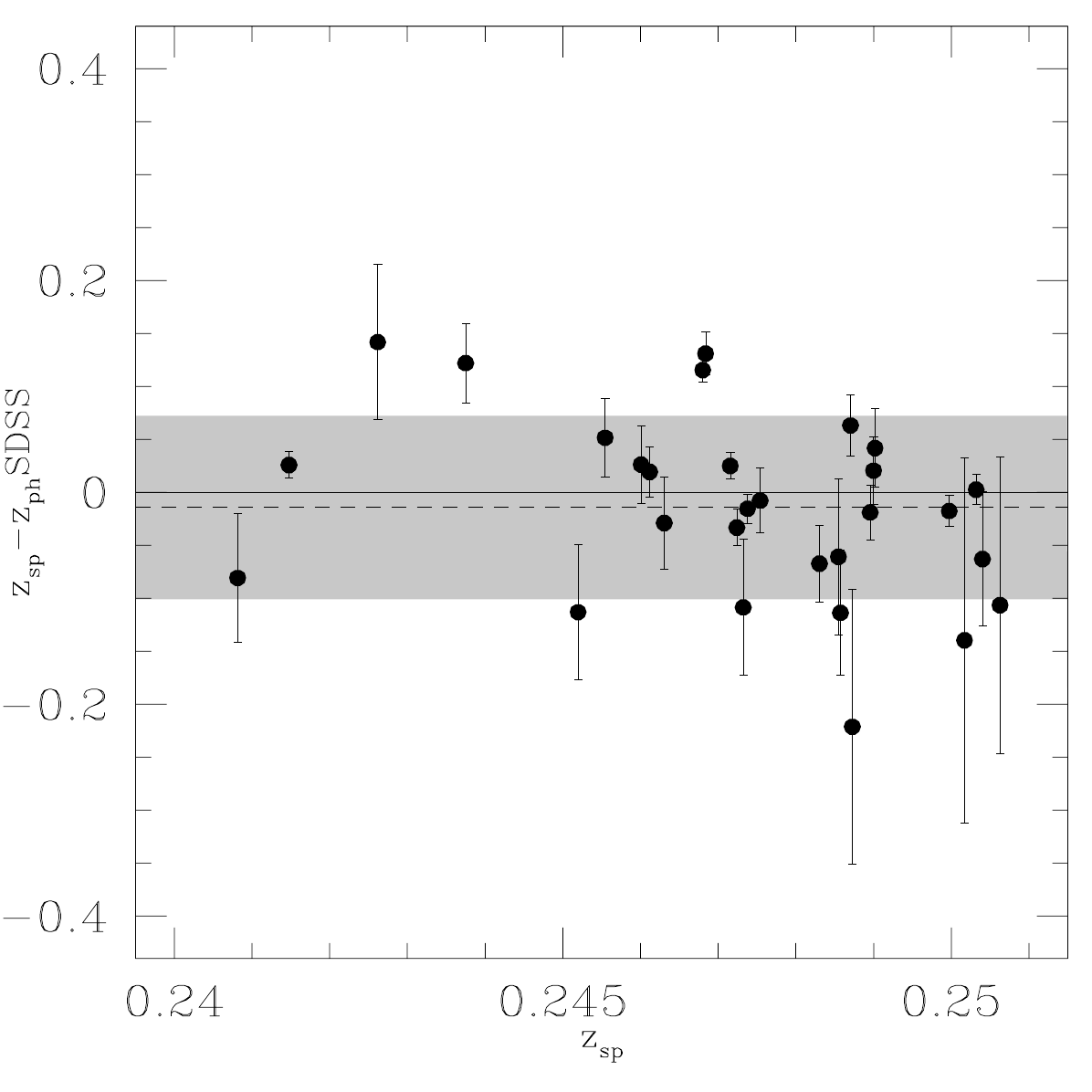}\\
\includegraphics[width=0.48\hsize]{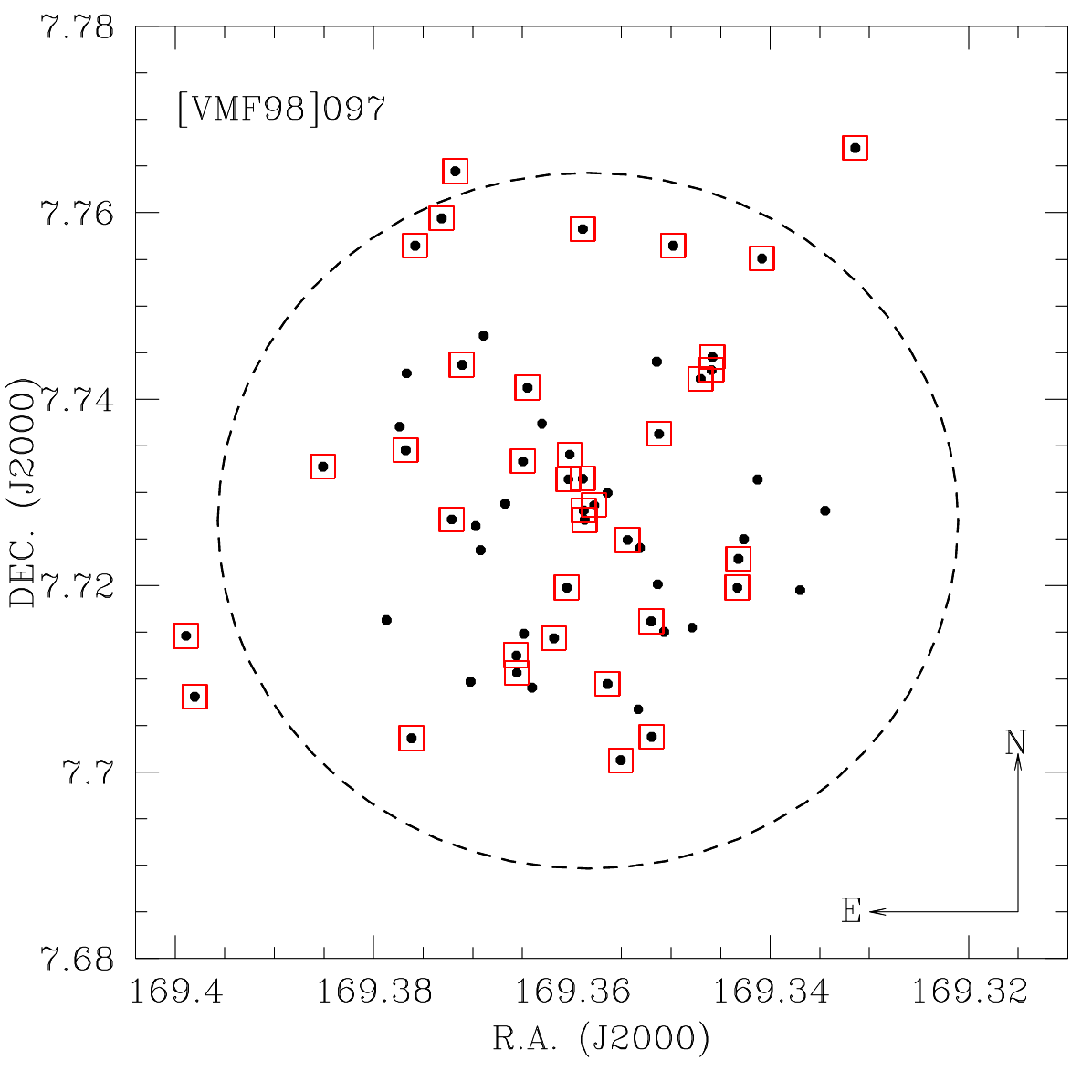}%
\includegraphics[width=0.48\hsize]{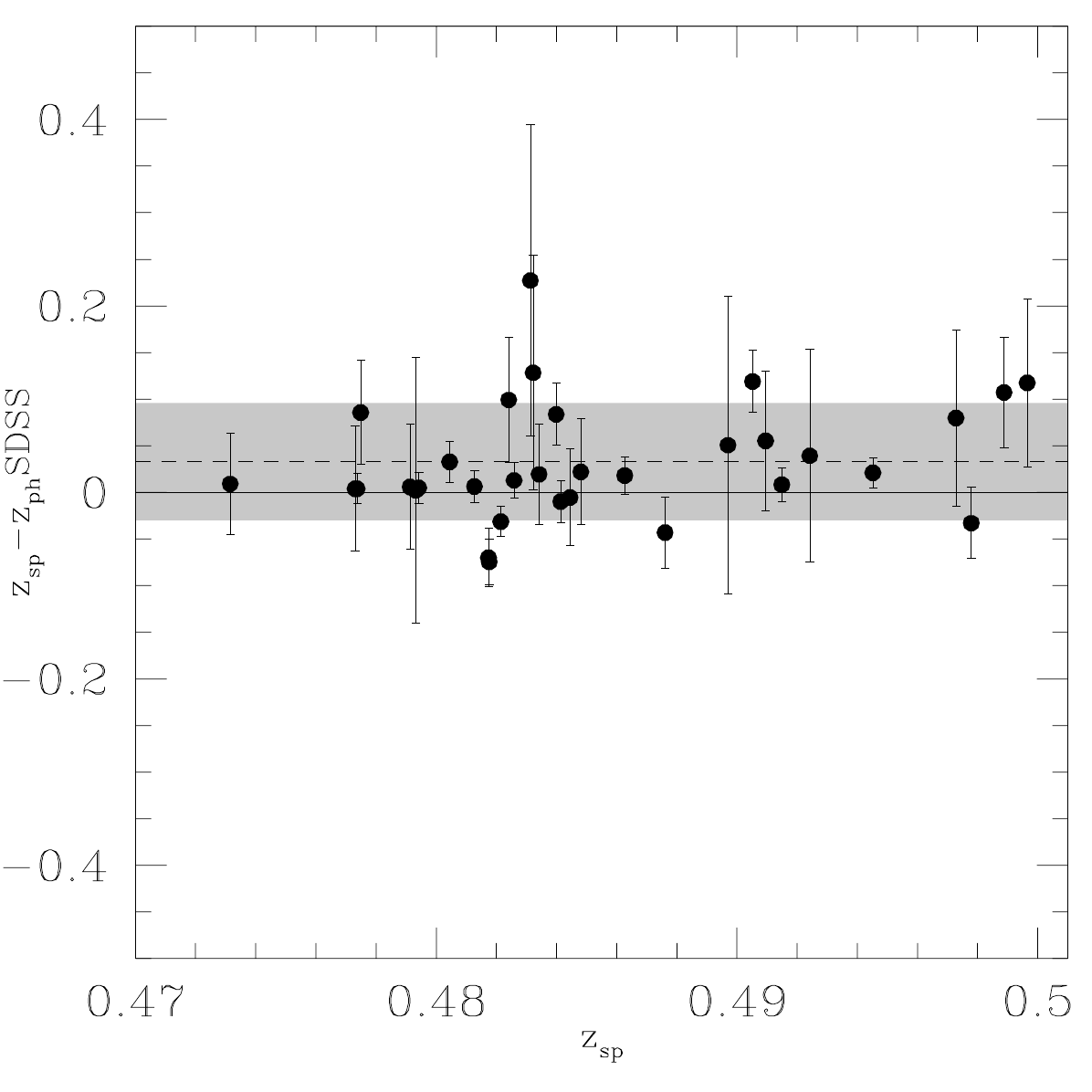}
\caption{Spectroscopic and photometric redshift comparison. 
Upper 
panels show the projected distribution of galaxies in the [VMF98]022 
field (left) and the
differences between spectroscopic and photometric redshift estimates vs spectroscopic values (right). In this panel, the mean value is represented by a dashed
line and the 1$\sigma$ 
scatter correspond to the grey region.  Lower panels are the same for 
galaxies in the cluster [VMF98]097.}
\label{compz0}
\end{centering}
\end{figure}

The photometric data observed with the Gemini telescopes were obtained in two 
passbands and for the reasons mentioned above determining photo-z was 
not possible with the use of the ANNz method.  For the galaxy
clusters: [VMF98]001, [VMF98]093, [VMF98]119 and [VMF98]148 without 
spectroscopy, 
we have used photometric redshifts extracted from the 
PHOTOZ tables (http://skyserver.sdss.org/CasJobs) of the SDSS-DR8.
These photometric redshifts use machine learning
techniques with training sets, similarly to those of ANNz. The SDSS
is 95\% complete for point sources up to $r^\prime \sim$
22.2 mag.  For the galaxy clusters in our sample with 
photometric redshifts from SDSS, the magnitude distributions are in agreement with 
the SDSS limiting magnitudes and completeness levels.
In order to estimate photometric redshift uncertainties, which are
significantly larger than spectroscopic measurements, we have made a 
comparison with two galaxy clusters:
[VMF98]022 and [VMF98]097 with both spectroscopic
and
photometric redshifts.  Figure~\ref{compz0} shows the
comparisons between our spectroscopic redshift estimates
with the photometric redshifts from the SDSS-DR8. In the left
panels, the projected
distribution of objects with spectroscopic (empty squares) and
photometric
(filled circles) redshifts are shown within the 0.75 Mpc region represented
by
the dashed circles.  The right panels correspond to the differences
z$_{sp}$ - z$_{ph}$ SDSS as a function of z$_{sp}$.  The mean of these
differences are
-0.014$\pm$ 0.087 for  [VMF98]022 and 0.033$\pm$ 0.063 for [VMF98]097,
which are
comparable to the uncertainties obtained by O'Mill et al. (2012).  These 
mean values are represented with dashed lines and the 1$\sigma$ 
scatter by the grey
region.

Since it is also seen that for $\Delta$V = 6000 km/s, contamination is less than 10\% with a still large number of true members, we have adopted this galaxy photometric redshift difference with respect to the cluster spectroscopic redshift in order to assess membership.

\subsection {Membership summary}

Cluster members are then, the 
galaxies within a projected radius of  0.75 Mpc from the X-ray emission
peak and  with spectroscopic $\Delta$V = 1$\sigma$ or photometric 
$\Delta$V $\sim$ 6000 km~s$^{-1}$.  Mean values of the cluster redshift with 
the members were obtained and the comparison with Mullis et al. (2003) 
gives mean differences of about 0.002 $\pm$ 0.005.
Table~\ref{table4} shows the final membership summary with the number of 
members assigned to the clusters in column 2, our mean redshift estimates in
column 3 and the redshift source in column 4.  
Our spectroscopic measurements
are identified as z$_{sp}$ while our 
photometric estimates with z$_{ph}$ and the SDSS estimates with z$_{ph}$ SDSS.
As mentioned above, the galaxy members of the clusters [VMF098]011 and
[VMF098]211 were not
possible to obtain because they have been observed in only two passbands.
Also the clusters  [VMF098]024 and  [VMF098]025 have similar redshifts with 
some members in common (see Figure~\ref{fov1}). 

\begin{deluxetable}{ccccc}
\tabletypesize{\scriptsize}
\tablewidth{0pc}
\tablecolumns{5}
\tablecaption{Member assignment for the sample of low X-ray galaxy clusters.\label{table4}}
\tablehead{
\colhead{[VMF98]}& \colhead{Number of}  & \colhead{our mean redshift} & \colhead{redshift} \\
\colhead{Id.} & \colhead{cluster members} & \colhead{} & \colhead{source}}
\startdata
001  & 22 & 0.495 	& z$_{ph}$ SDSS 	\\
004  & 32 & 0.292 	& z$_{ph}$   		\\
011  & -- & -- 	& -- 			\\
015  & 44 & 0.249 	& z$_{ph}$  		\\
022  & 26 & 0.247 	& z$_{sp}$ $|$ z$_{ph}$ SDSS   	\\
024  & 51 & 0.168 	& z$_{ph}$   		\\
025  & 46 & 0.163 	& z$_{ph}$   		\\
045  & 29 & 0.292 	& z$_{ph}$   		\\
046  & 38 & 0.200 	& z$_{ph}$  		\\
093  & 14 & 0.357 	& z$_{ph}$ SDSS  	\\
097 & 37 &  0.482 	& z$_{sp}$ $|$ z$_{ph}$ SDSS		\\
102 & 22 &  0.409 	& z$_{sp}$		\\
113 & 35 &  0.258 	& z$_{ph}$    		\\
119 & 20 &  0.692 	& z$_{ph}$ SDSS  	\\
124 & 12 &  0.185 	& z$_{sp}$ 		\\
148 & 16 &  0.697 	& z$_{ph}$ SDSS 	\\
211 & -- &  -- 	& --  			\\
214 & 41 &  0.198 	& z$_{ph}$  		\\
215 & 53 &  0.194 	& z$_{ph}$  		\\
\enddata
\end{deluxetable}

\section{Final comments and future plans}

Our project is centered on the study of low X-ray luminosity 
clusters of galaxies
at intermediate redshifts and the analysis of 
the morphological galaxy content. At the end of the project, the 
photometric and  spectroscopic
data will be available at http://astro.userena.cl/science/LowXrayClusters/.

This paper presents our sample and main goals. 
Nineteen galaxy clusters were selected with X-ray luminosities of 
L$_X \sim$ 0.5--45 
$\times$ $10^{43}$ erg s$^{-1}$ in the redshift range of 0.16 to 0.70, which were
observed at Las Campanas Observatory, Cerro Tololo Interamerican Observatory, 
and Gemini Observatory with different instruments and passbands.
We extensively discussed the photometric and spectroscopic observations and
the data reduction, which includes the galaxy identification and
cluster membership together with the  spectroscopic and photometric redshifts
and their error estimates. 

The second paper of the series (Nilo Castell\'on et al. 2014) have considered
optical properties and morphological content of the seven galaxy clusters observed 
with Gemini North and South telescopes
at 0.18 $<$ z $<$ 0.70.  The main results are an increment of the blue galaxy
fraction and a reduction of the lenticular fraction with redshifts.  The 
early-type 
fraction remains almost constant in the whole redshift range.  These
results are in agreement with those observed for massive clusters.

The third paper of the series (Gonzalez et al. 2015) have presented the weak 
lensing analysis of the galaxy clusters observed with Gemini telescopes.
We have determined the masses of seven galaxy clusters, six of them measured 
for the first time. The weak lensing mass determinations correlate with 
the X-ray luminosities following the observed M $-−$ L$_X$ relation.

In forthcoming papers, will be presented several analysis of the data, such us
the galaxy Luminosity Function, the RCS, density profiles and
morphological content.   In Valotto et al. (2016) we will take advantage
of the photometric data and cluster membership to study the Luminosity Function
of the galaxy cluster sample with 
different X-ray luminosities in the 
redshift range of 0.18 to 0.70.  Also, Alonso et al. (2016) will present an
analysis of 
the CMDs, the RCS, color-color diagrams and density profiles
for the galaxy clusters observed at LCO and CTIO in the redshift range of 
0.16 to 
0.30.  Finally, in Cuevas et al. (2016) we will study the morphological evolution
taking into account the nineteen galaxy clusters in the sample.  

A second part of this project includes more
spectroscopic measurements, which will allow us to search for 
substructures and the global cluster dynamics. The combination of photometric 
and spectroscopic data analysis
could provide useful hints to trace the evolutionary scenario of 
these low X-ray galaxy clusters.  
\vskip 0.5cm

\acknowledgments

Jos\'e Luis Nilo Castell\'on, Marcelo Jaque and Felipe Ramos acknowledge the financial support from Consejo Nacional de 
Investigaciones Cient\'ificas y T\'ecnicas de la Rep\'ublica Argentina (CONICET).  This work was partially supported by
grants by CONICET, Secretar\'ia de Ciencia y T\'ecnica (Secyt) of Universidad Nacional de C\'ordoba and
Ministerio de Ciencia y Tecnolog\'ia (MINCyT, C\'ordoba). JLNC  also acknowledges financial support from the Direcci\'on 
de Investigaci\'on de la Universidad de La Serena (DIULS), the Programa de Incentivo a la 
Investigaci\'on Acad\'emica, Programa DIULS de Iniciaci\'on Cient\'i­fica PI15142, and from ALMA-CONICYT postdoctoral 
program  No. 31120026. HC acknowledges support by Direcci\'on de Investigaci\'on de la Universidad de La Serena, PR13143.

SDSS-III is managed by the Astrophysical Research Consortium for the Participating Institutions of the SDSS-III Collaboration including the University of Arizona, the Brazilian Participation Group, Brookhaven National Laboratory, University of Cambridge, University of Florida, the French Participation Group, the German Participation Group, the Instituto de Astrof\'isica de Canarias, the Michigan State/Notre Dame/JINA Participation Group, Johns Hopkins University, Lawrence Berkeley National Laboratory, Max Planck Institute for Astrophysics, New Mexico State University, New York University, Ohio State University, Pennsylvania State University, University of Portsmouth, Princeton University, the Spanish Participation Group, University of Tokyo, University of Utah, Vanderbilt University, University of Virginia, University of Washington, and Yale University.


\end{document}